\documentclass[article]{aa}  
\usepackage{graphicx}
\usepackage{txfonts}
\usepackage[english]{babel}
\usepackage{amsmath}
\usepackage{amsfonts}
\usepackage{amssymb}
\usepackage{makeidx}
\usepackage{graphicx}
\usepackage{color}
\usepackage{appendix}
\usepackage{multirow}

\newcommand{\derd}{{\rm d}}
\newcommand{\Log}{{\rm Log}}
\newcommand{\Ms}{{\rm M}_\odot}

\newcommand{\au}{{\rm AU}}
\newcommand{\gw}{{\rm GW}}
\newcommand{\gc}{{\rm GC}}

\newcommand{\ibh}{{\rm IMBH}}

\newcommand{\imri}{{\rm IMRI}}
\newcommand{\inn}{{\rm in}}

\newcommand{\bh}{{\rm BH}}

\newcommand{\ARGdf}{\texttt{ARGdf~}}
\newcommand{\ARCHAIN}{\texttt{ARCHAIN~}}

\newcommand{\kms}{~{\rm km~s}^{-1}}

\begin{document}

\title{Merging stellar and intermediate-mass black holes in dense clusters: implications for LIGO, LISA and the next generation of gravitational wave detectors}


\titlerunning{Formation of IMRIs in Milky Way globular clusters and the Local Volume}	
\authorrunning{Arca Sedda, Amaro-Seoane, Chen}
\author{Manuel Arca Sedda \inst{1} \and Pau Amaro Seoane \inst{2,}\inst{3,}\inst{4,}\inst{5,}\inst{6} \and Xian Chen \inst{7,}\inst{4} }

\institute{Astronomisches Rechen-Institut. Zentrum f\"{u}r Astronomie der Universit\"{a}t Heidelberg, M\"onchhofstrasse 12-14, Heidelberg, D-69120, DE              
         \and         
			  Universitat Polit{\`e}cnica de Val{\`e}ncia, IGIC, 46022 Val{\`e}ncia, Spain
		\and DESY, Zeuthen, Germany
		\and         
			  Kavli Institute for Astronomy and Astrophysics at Peking University, Beijing 100871, P.R. China
		\and	  
			  Institute of Applied Mathematics, Academy of Mathematics and Systems Science, CAS, Beijing 100190, China
		\and
			  Zentrum f{\"u}r Astronomie und Astrophysik, TU Berlin, Hardenbergstra{\ss}e 36, 10623 Berlin, Germany
		\and
			Astronomy Department, School of Physics, Peking University, Beijing 100871, P.R. China	             
\\		\email{m.arcasedda@gmail.com}			
\\    	\email{amaro@upv.es}
\\		\email{xian.chen@pku.edu.cn }
}

\date{Received...; accepted...}

\abstract 
{The next generation of gravitational wave (GW) observatories would enable the detection of intermediate-mass black holes (IMBHs), an elusive type of black holes that are expected to lurk in the centre of massive clusters, dwarf galaxies and, possibly, AGN accretion discs. Intermediate mass ratio inspirals (IMRIs), composed of an IMBH and a compact stellar object, constitute one promising source of GWs audible to these detectors.} 
{We study the formation and evolution of IMRIs triggered by the interactions between two stellar BHs and an IMBH inhabiting the centre of a dense star cluster, with the aim of placing constraints on IMRIs formation rate and detectability.}
{We exploit direct $N$-body models varying the IMBH mass, the stellar BH mass spectrum, and the star cluster properties. Our simulations take into account the host cluster gravitational field and General Relativistic effects via Post-Newtonian terms up to order 2.5. These simulations are coupled with a semi-analytic procedure to characterise the evolution of the remnant IMBH after the IMRI phase.} 
{Generally, the IMRIs formation probability attains values $\sim 5-50\%$, with larger values corresponding to larger IMBH masses. Merging IMRIs tend to map out the stellar BH mass spectrum, thus suggesting that IMRIs could be used to unravel the role of dynamics in shaping BH populations in star clusters harboring an IMBH. After the IMRI phase, an IMBH initially nearly maximal(almost non-rotating) tend to significantly decrease(increase) its spin. Under the assumption that IMBHs grow mostly via repeated IMRIs, we show that only IMBH seeds sufficiently massive ($M_{\rm seed} > 300\Ms$) can grow up to $M_{\ibh} >10^3\Ms$ in dense globular clusters. Assuming that these seeds form at a redshift $z\sim 2-6$, we find that around $1-5\%$ of them would reach typical masses $\sim 500-1500\Ms$ at redshift $z=0$ and would exhibit low-spins, generally $S_\ibh < 0.2$. Measuring the mass and spin of IMBHs involved in IMRIs could help unravelling their formation mechanisms. We show that LISA can  detect IMBHs in Milky Way globular clusters with a signal-to-noise ratio SNR$=10-100$, or in the Large Magellanic Cloud, for which we get an SNR$=8-40$. More in general, we provide the IMRIs merger rate for different detectors, namely LIGO ($\Gamma_{\rm LIGO} = 0.003-1.6$ yr$^{-1}$), LISA ($\Gamma_{\rm LISA} = 0.02-60$ yr$^{-1}$), ET ($\Gamma_{\rm ET} = 1-600$ yr$^{-1}$), and DECIGO ($\Gamma_{\rm DECIGO} = 6-3000$ yr$^{-1}$).}
{Our simulations explore one possible channel for IMBH growth, i.e. via merging with stellar BHs in dense clusters. We have found that the mass and spin of IMRIs'  components and the merger remnant  encode crucial insights on the mechanisms that regulate IMBH formation. Our analysis suggests that the future synergy among GW detectors would enable us to fully unveil IMBHs formation and evolution.}


\maketitle
%

\section{Introduction}

Intermediate mass black holes (IMBH), with masses in the range $10^2-10^5\Ms$, might represent the missing link between stellar and supermassive BHs (SMBHs). Dense stellar systems, such as globular clusters (GCs), are thought to be ideal factories for the formation of IMBHs, either via the collapse of a very massive star assembled through stellar collisions \citep{zwart02, freitag06b,freitag06c, giersz15, mapelli16}, or via multiple interactions and mergers between stars and stellar-mass BHs \citep{giersz15, dicarlo19, rizzuto21, gonzalez21}. 
Aside from the scenarios above, further formation mechanisms for IMBHs include: direct collapse of massive stars with an extremely low metallicity \citep{madau01,bromm02,ohkubo09,spera17} or of gaseous clouds in the early Universe \citep{latif13}, IMBH seeding in high redshift, metal poor, galactic halos \citep{bellovary11}, IMBH formation in satellite galaxies' nuclei later accreted in their host galaxy halo \citep{bellovary10}, or in AGN accretion discs \citep{mckernan12} and in galactic discs circumnuclear regions \citep{taniguchi00}.

Several processes can mimic an IMBH in GCs, like anisotropies in the cluster kinematics \citep{zocchi}, or the presence of a dense subsytem of stellar mass BHs harbored in the cluster centre \citep{vandermarel10,AS16,AAG18b,weatherford18}. Nevertheless, a few observational IMBH candidates have been found in Galactic GCs
\citep{noyola10,lu13,lanzoni13,kiziltan17}, whose masses and half-mass radius could be connected with the host cluster observables \citep{Baumgardt17,AAG18a,AAG19}. Therefore, finding an unique way to unravel the presence of IMBHs in GCs represents one of the most interesting challenges in modern astronomy \citep[for recent reviews see][]{mezcua17,greene19}. 

Despite that a striking observational evidence for the existence of IMBHs with masses above $10^3\Ms$ is still missing, the detection of GW190521, a gravitational wave (GW) source associated with the merger between two BHs with masses $66\Ms$ and $85\Ms$ \citep{gw190521a}, marks the discovery of the first IMBH with a confirmed mass $>100\Ms$. 
Detecting IMBHs via GW emission represents an appealing possibility from the perspective of the next generation of GW observatories. A compact object orbiting the IMBH can enter the regime dominated by GW emission and emit low-frequency GWs \citep[][Arca Sedda et al in prep.]{konstantinidis13,leigh14,haster16,macleod16,rizzuto21}, making systems like this a  promising class of sources -- denominated intermediate-mass ratio inspirals (IMRIs) -- audible to future detectors like the laser interferometer space antenna \citep[ LISA][]{Will04,seoane07,amaro12,seoane18}.

However, in the highly dense regions that characterise star clusters centres, the formation of IMRIs is not a smooth process. Indeed, due to the continuous interactions with stars, an IMRI ``progenitor'', namely a tight IMBH-BH binary, might be subjected to strong perturbation induced, for instance by a passing-by BH. The three-body interaction involving the IMBH and the two BHs can lead to a variety of end states, including the formation of an IMRI, a stellar BH binary, the ejection of one BH, or even both, or the development of a head-on collision. At some extent, this scenario is similar to what is expected to happen in galactic nuclei, where an SMBH can capture a compact object to form an extreme mass ratio inspiral. However, in the case of IMRIs the picture is complicated by the fact that this chaotic process can transfer to the IMBH an amount of energy sufficient to displace it sensibly from the cluster centre. Differently from galactic nuclei, where the SMBH remains well seated in the galactic potential well, the IMBH motion makes hard the use of any analytical approach to solve IMRIs dynamics. Although tight IMBH-BH binaries can temporarily form in the centre of clusters with IMBH mass $\sim 100-1000\Ms$ \citep{konstantinidis13,macleod16} and could last up to $10^7$ yr \citep{macleod16}, it is extremely difficult to predict the actual amount of BHs that, at any time, interact with the IMBH. The few studies in the literature assume that the IMBH is already at the centre of the host cluster when stellar BHs form and sink toward the cluster centre \citep{leigh14,haster16}, thus they potentially neglect the important phase during which the BH reservoir is depleted by BH-BH interactions. Recent models in which the IMBH formation is a byproduct of stellar evolution and collisions pointed out that a sizeable number of BHs could be still present during the earliest phases of IMBH seeding, when its mass is $\lesssim 500\Ms$ and the cluster age is $<0.01-1 Gyr$ \citep[see e.g.][]{dicarlo19,rizzuto21,gonzalez21}, but the number of BHs reduces to a few over longer timescales, when the IMBH fully grew to $>10^3\Ms$ \citep{giersz15,AAG19}.

Quantifying the branching ratios for IMRIs formation mechanisms  constitute a fundamental step to assess the probability to observe these GW sources with the next generation of space-based detectors like LISA\footnote{\url{https://www.elisascience.org/}} \citep{Will04,seoane07,amaro12}, TianQin \citep{tianqin16} or Taiji \citep{taiji17}.

In this paper, we model the  formation of an IMRI mediated by the interaction between an IMBH and two stellar mass BHs, aiming at unravelling the role of the environment, the BH natal spin, and the IMBH mass in determining IMRIs properties.  

To reach the aim, we use $N$-body simulations that take into account in particles' equations of motion both the star cluster gravitational potential and post-Newtonian corrections at 1, 2, and 2.5 order \citep{mikkola08,ASCD19}. Varying the IMBH and BHs masses, their orbital configuration, and the host cluster structural properties, we build-up 7 sets consisting of 4,000 simulations each.

The paper is organized as follows: in Section \ref{num} we present and summarise the numerical setup used to model the IMBH-BH-BH interaction, in Section \ref{res} we present and discuss the main results of our simulations and the implications for IMBH formation and evolution, Section \ref{gra} focuses on the implications for GW astronomy, while Section \ref{end} is devoted summarise our main findings.

\section{Initial conditions} 
\label{num}

\begin{figure}
    \centering
    \includegraphics[width=7cm]{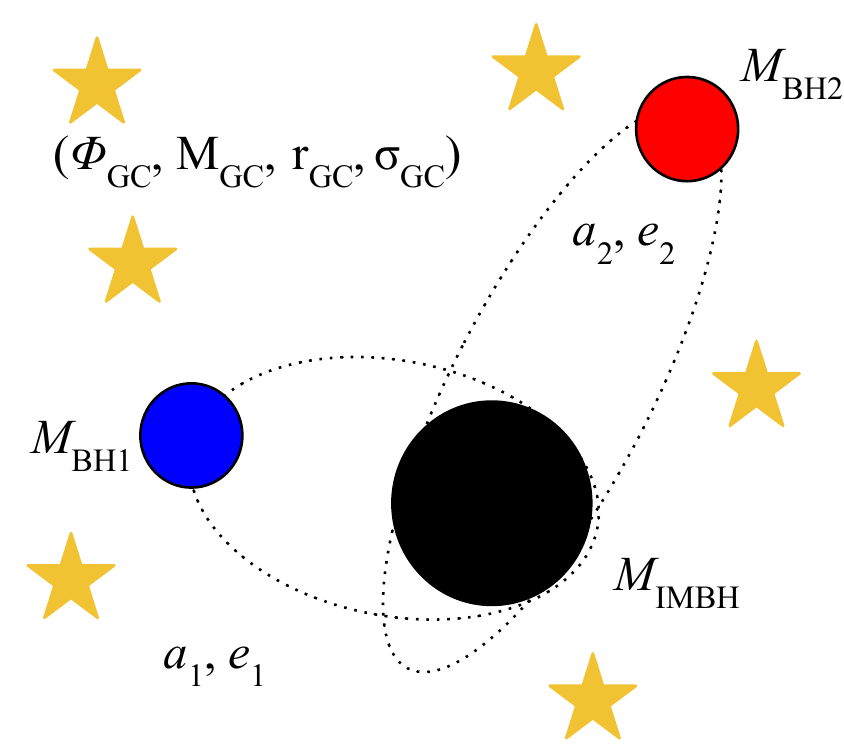}
    \caption{Sketch of the IMBH-BH-BH triple configuration. We mark the main quantities that characterise the triple -- IMBH mass ($M_\ibh$), stellar BH masses ($M_{\bh 1,2}$, orbital semimajor axis and eccentricity of the IMBH-BH1 ($a_1,~e_1$) and IMBH-BH2 ($a_2,~e_2$) systems -- and the host cluster -- gravitational potential ($\Phi_\gc$), total mass ($M_\gc$), typical radius ($r_\gc$), and velocity dispersion ($\sigma_\gc$).}     
\label{f1}
\end{figure}

We simulate the evolution of the innermost region of the cluster, modelling the IMBH and two stellar mass BHs as live particles and the remaining cluster as an external static potential. In order to explore the parameter space, we create 7 different models, each one consisting of 4,000 simulations gathered in four sub-classes depending on the IMBH mass, for a total of 28,000 simulations. Our model is sketched in Figure \ref{f1}. 

We adopt four values of the IMBH mass, namely ${\rm Log}(M_\ibh/\Ms) = 2,~3,~4,~5$. This range covers typical values of putative IMBH masses forming in stellar systems of various sizes, from young and open clusters, to globular clusters, and up to nuclear clusters. We note that the lowest value taken for $M_\ibh$ will likely involve mergers with mass ratio $> 0.1$, thus they fall outside the range of IMRIs. Nonetheless, exploring the low end of IMBH mass function will help us in better understanding the perspectives of IMBH-BH mergers from the point of view of both low- \citep[e.g.][]{Will04,seoane18} and high-frequency \citep[e.g.][]{MandelEtAl2008,gair11,abbott17f} GW detectors, especially in the light of the recent discovery of GW190521, a GW source associated with the formation of an IMBH with mass $142\Ms$ \citep{gw190521a}.

The properties of the cluster in which the IMBH is embedded, which define the cluster potential, are varied depending on the model. The cluster density profile is assumed to be either a Dehnen sphere with inner slope $\gamma_\gc = 0.5$ (model S0, S3, S4, S5, S6) or $1.0$ (model S1), or a Plummer sphere (model S2). In both cases, the cluster half-light radius is assumed to be $R_{\rm eff} = 3.4$ pc, namely the mean value of Milky Way globular clusters \citep{harris14}. The mass of the cluster is calculated via the scaling provided by \cite{AS16}, connecting the host cluster mass $M_\gc$ with the total ``dark'' mass, inhabiting the cluster's centre, comprised of either an IMBH or a sizable population of stellar BHs
\begin{equation}
\Log \left(\frac{M_\ibh}{\Ms}\right) = \alpha \Log \left(\frac{M_\gc}{\Ms}\right) - \beta.
\label{mcmibh}
\end{equation}
with $\alpha = 0.999 \pm 0.001$ and $\beta = 2.23 \pm 0.009$. The cluster typical radius is thus calculated from the assumed $R_{\rm eff}$ and the adopted mass profile. Upon these assumptions an IMBH with mass $M_\ibh = 10^2(10^5)\Ms$ is associated with a cluster mass of $ M_\gc = 1.7\times 10^4(1.7\times 10^7)\Ms$, thus our models ideally span the mass range of young clusters, globular clusters, nuclear clusters, and dwarf galaxies nuclei. We assume that the IMBH is orbited by two stellar BHs, since the IMBH is expected to be the most massive object in the cluster and the dominant element in determining the dynamics. The choice of limiting the number of BH companions to two is dictated by the numerical evidence that IMBHs tend to form after the reservoir of BH population diminished severely through dynamical interactions \citep{zwart02,giersz15}. Compared to other works focused on a similar topics, our models do not rely on any  preferential configuration or initial hierarchy for the IMBH-BH-BH system, since we expect that the evolution around the IMBH will be mostly driven by the chaotic interactions involving the compact objects surrounding the IMBH \citep[e.g.][]{konstantinidis13}.

We assume that the two BHs move on Keplerian orbits around the IMBH and that the centre of mass of the three BHs coincides with the cluster centre. Note that this choice ensures that the lower mass IMBH are not necessarily in the cluster centre at the beginning of the simulation. For each IMBH-BH orbit, we draw the orbital eccentricity $e_{1,2}$ from a thermal distribution $P(e){\rm d} e = 2e{\rm d}e$ \citep{Plum}. The semimajor axes of the two orbits are selected either from a distribution flat in logarithmic values limited between $a_{1,2} = 0.1-10^4$ AU (set S5) or according to the overal cluster mass distribution (S0-4 and S6) which for a Dehnen model is given by:
\begin{equation}
M(r) = M_{\rm GC} \left(\frac{r}{r+r_{\rm GC}}\right)^{3-\gamma_{\rm GC}},
\end{equation}
where $r_{\rm GC}= 4/3 R_{\rm eff}  \left(2^{1/(3-\gamma)}-1\right)$ is the cluster scale radius \citep{Deh93}. In the latter case, the maximum semimajor axis allowed is given by the distance to the cluster centre at which the cluster mass is $M(r) = 60\Ms$, i.e. twice the typical stellar BH mass \citep{spera17}, thus inverting the equation above we get $a_{\rm max} \equiv r(60\Ms)$. This choice implies that increasing the cluster mass at fixed $R_{\rm eff}$ leads to a smaller value of $a_{\rm max}$, e.g. for $M_{\rm IMBH} = 100(10^5)\Ms$ we obtain $a_{\rm max} = 20,000 ( 200 ) $ AU. 
Note that the assumption on the semimajor axis distribution in S0-4 and S6 ensures that the initial position of the BHs follows the underlying mass distribution of the host cluster. As opposed to this, the assumption of a logarithmically flat $a_{1,2}$ distribution for S5 permits us to explore the role of the semimajor axis in determing IMRIs formation.

The BH mass spectrum is also allowed to vary: we use the BH mass spectrum derived by \citep[SM17][]{spera17}, assuming a progenitor metallicity of $Z = 0.0002$ for models S0, S1, S2, and S5 and $Z = 0.02$ for model S6; a power-law mass spectrum with slope $2.2$ \citep[O16][]{oleary16} in the range $3-30 \Ms$ (model S3); or a flat distribution in the range $3-30 \Ms$ (hereafter FLAT, model S4). The three different BH mass spectra adopted describe two possible situations, one in which the BH population reflects the original population of stars and another in which dynamics operated a selection on the BH mass distribution either mild (power-law distribution) or sufficiently strong to erase any memory of the original mass function (flat distribution).

\begin{table*}
\caption{Main properties of our models}
\begin{center}
\begin{tabular}{cccccccccccc}
\hline
\hline
ID & $M_\ibh$ & $M_\gc$ & $\rho_\gc$ & $\gamma_\gc$ & $a$ &$e$ &BH & $m_{\bh, min/max}$ & $Z$ & $N_{\rm sim}$ \\ 
   & $\Ms$    & $\Ms$   &            &              &     &    &   & $\Ms$              &     & \\
\hline
S0 & $10^2-10^5$ & $1.7\times 10^4 - 1.7\times 10^7$ & Dehnen & $0.5$ & density & thermal & SM17 & $3.3-53.4$ & 0.0002 & 1,000 $\times$ 4 \\
S1 & $10^2-10^5$ & $1.7\times 10^4 - 1.7\times 10^7$ & Dehnen & $1.0$ & density & thermal & SM17 & $3.3-53.4$ & 0.0002 & 1,000 $\times$ 4 \\
S2 & $10^2-10^5$ & $1.7\times 10^4 - 1.7\times 10^7$ & Plummer& $0.0$ & density & thermal & SM17 & $3.3-53.4$ & 0.0002 & 1,000 $\times$ 4 \\
S3 & $10^2-10^5$ & $1.7\times 10^4 - 1.7\times 10^7$ & Dehnen & $0.5$ & density & thermal & O+16 & $3-30$ & 0.0002 & 1,000 $\times$ 4 \\
S4 & $10^2-10^5$ & $1.7\times 10^4 - 1.7\times 10^7$ & Dehnen & $0.5$ & density & thermal & FLAT & $3-30$ & 0.0002 & 1,000 $\times$ 4 \\
S5 & $10^2-10^5$ & $1.7\times 10^4 - 1.7\times 10^7$ & Dehnen & $0.5$ & logflat & thermal & SM17 & $3.3-53.4$ & 0.0002 & 1,000 $\times$ 4 \\
S6 & $10^2-10^5$ & $1.7\times 10^4 - 1.7\times 10^7$ & Dehnen & $0.5$ & density & thermal & SM17 & $2.8-25$ & 0.02   & 1,000 $\times$ 4 \\
\hline
\end{tabular}
\tablefoot{Col 1: set ID. Col 2: IMBH mass range. Col 3: Cluster mass range. Col 4: cluster density profile adopted. Col 5: inner slope of the cluster density profile. Col 6-7: distribution adopted for the semimajor axis and eccentricity. Col 8: stellar BH mass spectrum adopted. Col 9: stellar BH mass range. Col 10: metallicity of BH progenitors. Col 11: number of simulations performed.}
\label{tabt1}
\end{center}
\end{table*}

All simulations are performed using \ARGdf \citep{ASCD19}, a modified version of the \ARCHAIN code that implements post-Newtonian (PN) dynamics and {\it algorithmic regularization} to handle close encounters and strong collisions \citep{mikkola99,mikkola08}. For our purposes, we include in our treatment only 1, 2, and 2.5 order PN terms. Additionally, \ARGdf allows the user to take into account the gravitational field generated by the host stellar system and a dynamical friction term in particles' equations of motion. Simulations are halted either if: one of the two BHs merge with the IMBH, the two BHs merge together, one of the BHs is ejected away, the simulated time exceeds $t = 1$ Gyr, or the runtime exceeded 1 hr. 
 
The choice of a maximum simulation time of $1$ Gyr is twofold. On the one hand, this is the typical timescale ({\it evaporation time}) over which the IMBH-BH systems might disrupt due to interactions with passing-by stars, as explained in the next section. On the other hand, this limit enables us to keep a good balance between the computational load and the data storage -- the data required $\sim 4$ Tb of storage space and around 2 month of computational time -- and the reliability of the models. Note that carrying out the simulations for a time $>10^9$ yr implies performing the BH orbit integration $>10^6$ times. Such long integration could lead the integration error to increase considerably, despite the \ARGdf code enables an accuracy over conserved quantities to a level $<10^{-12}$. A summary of all models considered is provided in Table \ref{tabt1}.

\section{Results}
\label{res}

\subsection{IMRIs formation and merger}
\label{sec:imris}

\begin{figure*}
    \centering
    \includegraphics[width=16cm]{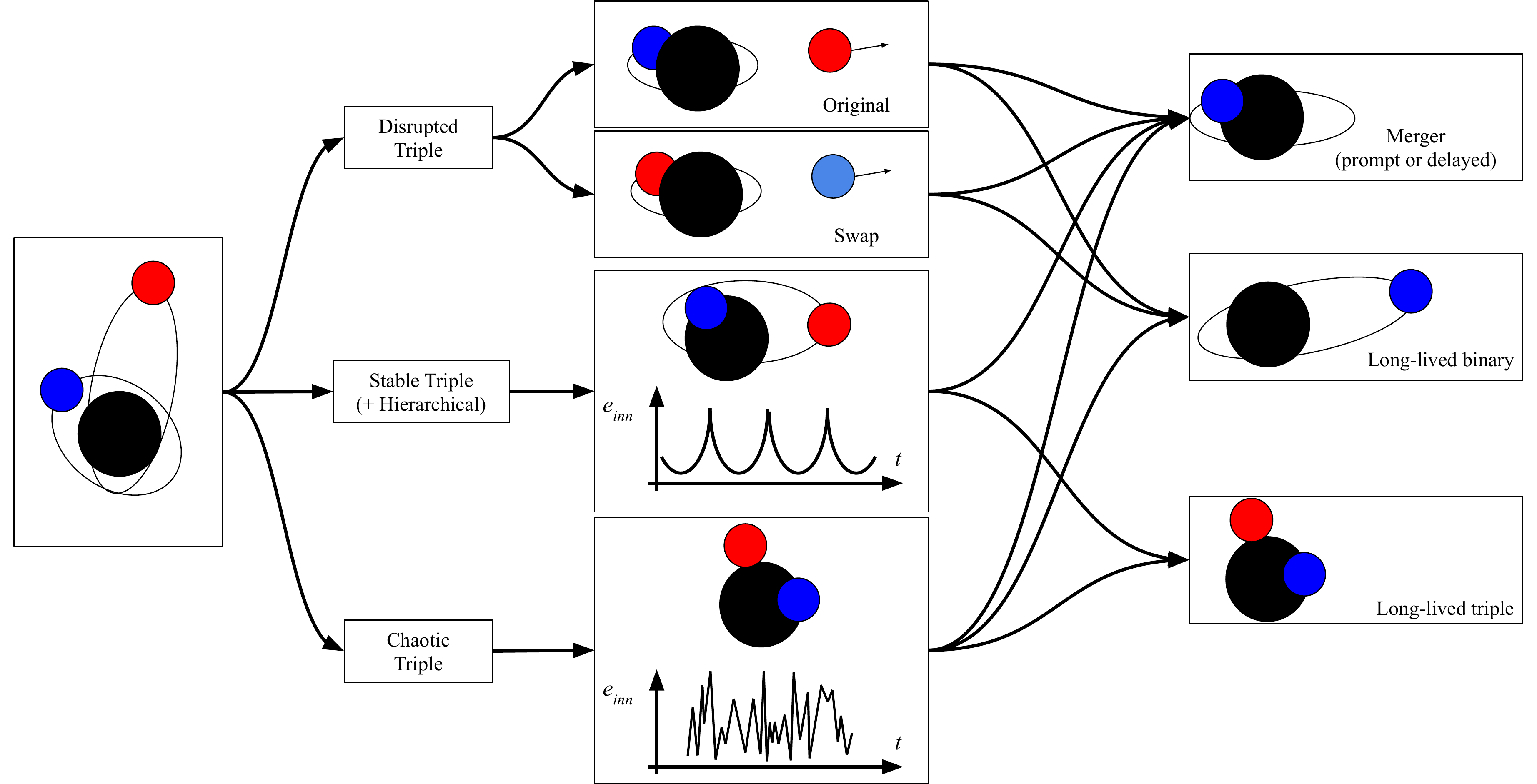}
    \caption{Schematic view of IMBH-BH-BH triple evolution.}
    \label{fig:f3}
\end{figure*}

\begin{table*}
\begin{center}
\caption{Main results of $N$-body simulations.}
\label{tab:mergers} 
\centering
\begin{tabular}{c|cccc|cccc}
\hline
\hline
ID   & $f_{\rm bnd}$ & $f_{\rm dis}$ & $f_{\rm IMRI}$ & $N_{\rm bbh}$ & \multicolumn{4}{c}{$f_{\rm IMRI,ibh}$} \\
     &  & & & & $10^2\Ms$ &$10^3\Ms$ &$10^4\Ms$ &$10^5\Ms$ \\
\hline
S0  & $ 0.23 $ & $ 0.49 $ & $ 0.28 $ & $ 0 $ & $ 0.03 $ & $ 0.27 $ & $ 0.51 $ & $ 0.30 $ \\
S1  & $ 0.17 $ & $ 0.61 $ & $ 0.22 $ & $ 2 $ & $ 0.03 $ & $ 0.27 $ & $ 0.38 $ & $ 0.16 $ \\
S2  & $ 0.38 $ & $ 0.37 $ & $ 0.25 $ & $ 0 $ & $ 0.00 $ & $ 0.12 $ & $ 0.44 $ & $ 0.44 $ \\
S3  & $ 0.26 $ & $ 0.42 $ & $ 0.32 $ & $ 2 $ & $ 0.04 $ & $ 0.48 $ & $ 0.44 $ & $ 0.31 $ \\
S4  & $ 0.24 $ & $ 0.46 $ & $ 0.30 $ & $ 0 $ & $ 0.03 $ & $ 0.36 $ & $ 0.48 $ & $ 0.30 $ \\
S5  & $ 0.26 $ & $ 0.31 $ & $ 0.42 $ & $ 22 $ & $ 0.21 $ & $ 0.29 $ & $ 0.31 $ & $ 0.46 $ \\
S6  & $ 0.27 $ & $ 0.39 $ & $ 0.34 $ & $ 1 $ & $ 0.03 $ & $ 0.49 $ & $ 0.49 $ & $ 0.33 $ \\
\hline
\end{tabular}
\tablefoot{Col 1: set ID. Col 2-4: fraction of models that remain bound, get disrupted, or form an IMRI. Col 5: number of mergers between stellar BHs. Col 6-9: fraction of models forming an IMRI for different values of the IMBH mass.}
\end{center}
\end{table*}

In this section we discuss the main results of our simulations. Hereafter we refer indifferently to IMBH-BH and IMRI although the smallest value of the IMBH mass adopted ($10^2\Ms$) leads to IMBH-BH mergers with a mass ratio larger than expected for IMRIs. 
The outcomes of our simulations can be classified in three main categories: 
\begin{itemize}
\item[a)] the IMBH-BH-BH system remains bound over the simulated time ({\it bound}, $f_{\rm bnd}$);
\item[b)] one of the stellar BHs is ejected away leaving behind an IMBH-BH binary ({\it disrupted}, $f_{\rm dis}$); 
\item[c)] one of the BH merges with the IMBH ({\it mergers}, $f_{\imri}$). 
\end{itemize}
Figure \ref{fig:f3} provides a simplistic sketch of the possible outcomes of our simulations. Table \ref{tab:mergers} shows the percentage of models falling in each of these categories for the different models explored. On average, we note that mergers constitute the $f_{\imri} \sim 20-32\%$ of models, with little dependence on the initial conditions assumed. 
Models falling in category a) or b) do not exclude automatically an IMBH-BH merger. In case a), i.e. a bound IMBH-BH-BH, the triplet can arrange in a configuration in which the IMBH forms a tighter bound with one of the BHs while the other orbits around their common centre of mass. In this case, the triple can either evolve chaotically or undergo secular effects like the so called Kozai-Lidov mechanism \citep{kozai62,lidov62} that can trigger the eccentricity of the innermost IMBH-BH binary to grow to values close to unity. 

In case b), i.e. ejection of one of the stellar BHs, the evolution of the remaining IMBH-BH binary will be due to the sum of two contributes, namely energy removal from binary-single interactions and GW emission. 

In both cases a) and b), binary-single interactions compromise the IMBH-BH survival over a typical {\it evaporation time} \citep{bt,stephan16,hoang18} 
\begin{align}
t_{\rm ev} =& \frac{\sqrt{3}\sigma_g}{32\sqrt{\pi}G\rho_g\ln\Lambda a} \frac{m}{m_*} = \nonumber \\
            & 1.3\times 10^{10} {\rm ~yr} 
            \left(\frac{\sigma_g}{5{\rm ~km~s^{-1}}}\right)
            \left(\frac{10^5{\rm ~M}_\odot ~pc^{-3}}{\rho_g}\right) \times \nonumber \\
            & \times \left(\frac{0.1{\rm ~AU}}{a}\right)\left(\frac{1030{\rm ~M}_\odot}{M_{\rm ibh}+M_{\bh}}\right)\left(\frac{m_*}{30{\rm ~M}_\odot}\right),
\label{eva}
\end{align}
where $m_*$ is the average stellar mass in the nucleus, $\rho_g$ is the stellar density, $\sigma_g$ is the cluster velocity dispersion, and $\ln\Lambda=6.5$ is the Coulomb logarithm. In our simulations, the initial evaporation time ranges between $t_{\rm ev} = 10^7-10^9$ yr, depending on the orbital properties, the IMBH mass, and the cluster structure.
If the IMBH-BH entered the IMRI phase, binary-single interactions are expected to play little to no effect on its evolution \citep{seoane18}. In this case, the IMRI will continuously shrink emitting GWs until coalescence, which takes place on a timescale \citep{peters64}
\begin{align}
t_\gw =&  \displaystyle \frac{5}{256}\frac{c^5 a_\inn^4 (1-e_\inn^2)^{7/2}}{G^3M_\ibh M_\bh(M_\ibh+M_\bh)} = \nonumber \\
       & 10^6{\rm ~yr}\left(\frac{a_\inn}{0.1\au}\right)^4\left(1-e_\inn^2\right)^{7/2}\times \nonumber \\ 
       & \times \left(\frac{10^3\Ms}{M_\ibh}\right) \left(\frac{30\Ms}{M_\bh}\right)\left(\frac{1030\Ms}{M_\ibh+M_\bh}\right).
\label{peters}
\end{align}

The long term effect of multiple perturbers onto the evolution of the IMBH-BH-BH system cannot be captured by our simulations, thus we decided to exclude from the analysis all models in which the three bodies remain bound by the end of the simulation. We thus consider only systems falling in cases b and c.
In simulations where one BH is ejected away (case b), we label the remaining IMBH-BH binary as merger if $t_\gw < t_{\rm ev}$, as these systems are likely to merge before dynamical encounters break them.

Table \ref{tab:mergers} summarises the main results of our simulations, highlighting the fraction of bound systems, disrupted systems, and mergers. As indicated in the table, we found a handful of models in which the two stellar BHs undergo merger, whose number is $N_{\rm bbh} < 1-22$. This effect is maximized in set S5 and for models with IMBH mass $M_{\ibh} = 100-1000\Ms$.

\begin{figure}
\includegraphics[width=\columnwidth]{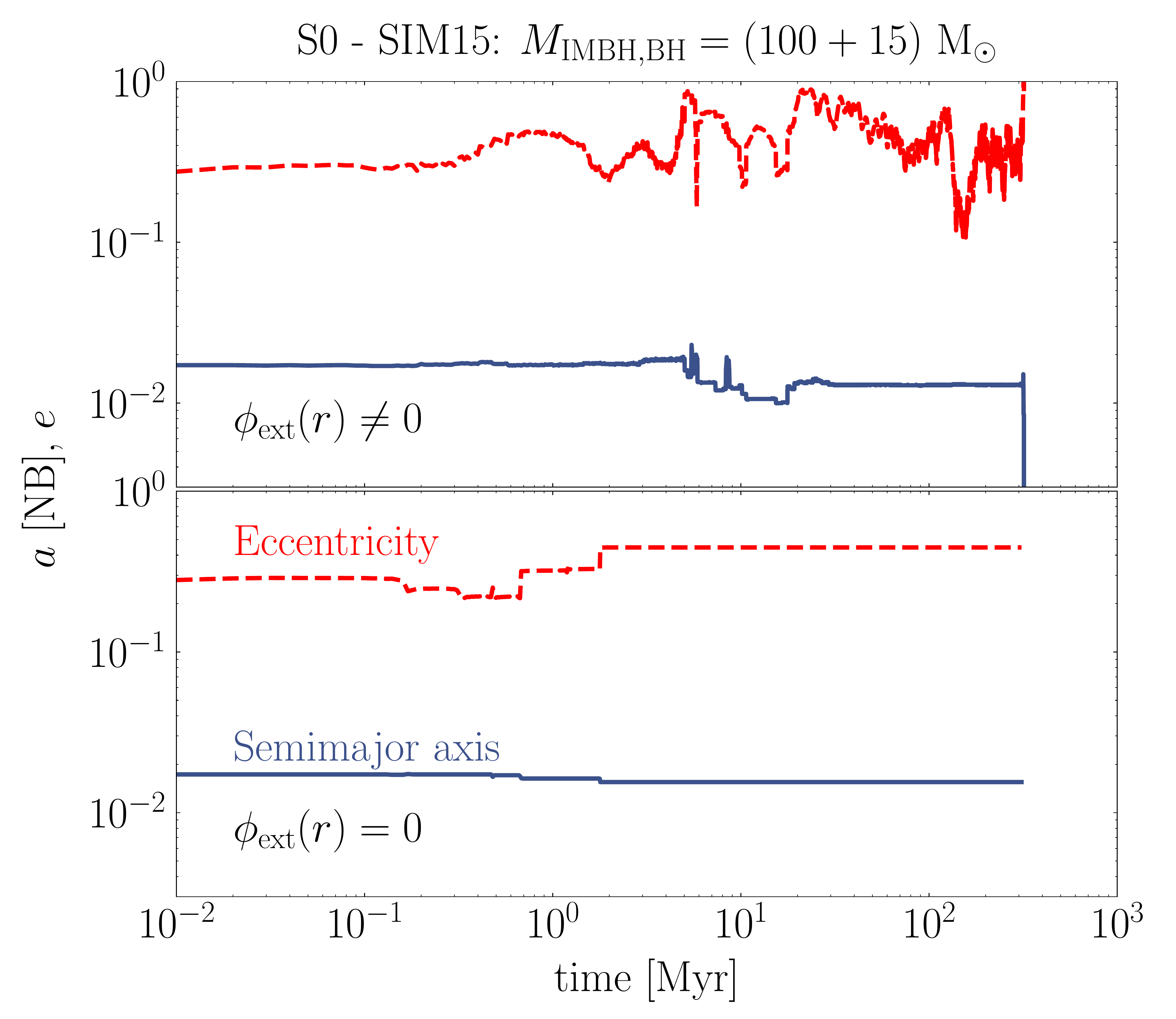}
    \caption{Time evolution of the semimajor axis (straight blue line) and eccentricity (dotted red line) for one of the simulations performed in S0. Panels show the case with (top panel) and without (bottom panel) the external potential of the cluster.}
	\label{fig:example}
\end{figure}

Clearly, while the categorization provided suggests three well separate classes, it must be noted that models falling in case c) (mergers) might have undergone a chaotic triple phase, or secular effects that triggered the IMBH-BH merger. Figure \ref{fig:example} shows one such example: an IMBH with mass $M_\ibh = 10^2\Ms$ forms a tight binary with a stellar BH with mass $M_\bh = 15\Ms$, the binary is subjected to perturbation of the outer BH with mass $M_\bh = 5\Ms$ that causes a continuous oscillation of the eccentricity up to the point (case b) -- around 300 Myr from the beginning of the simulation -- at which the eccentricity peaks at values $\sim 0.999998$, GW emission kicks in and start dominating the IMBH-BH binary evolution eventually culminating in a merger (case c). The bottom panel in Figure \ref{fig:example} shows the evolution of the same system in absence of the external potential. It can be seen that when the external potential is not accounted for in the simulation, the three BHs undergo a faster evolution that leads to the ejection of one BH on a timescale $<2$ Myr, leaving behind an IMBH-BH binary with a merger timescale $\sim 10^22$ yr. In this case, thus, the external potential facilitates the IMBH-BH merger by favouring a longer and more efficient interaction among the three BHs. Nonetheless, it must be noted that predicting the effect of an external potential onto the evolution of the three bodies is not trivial, as it does not necessarily facilitate the binary merger \citep[see e.g.][]{arca20b,petrovich17}.

\begin{figure}
\includegraphics[width=\columnwidth]{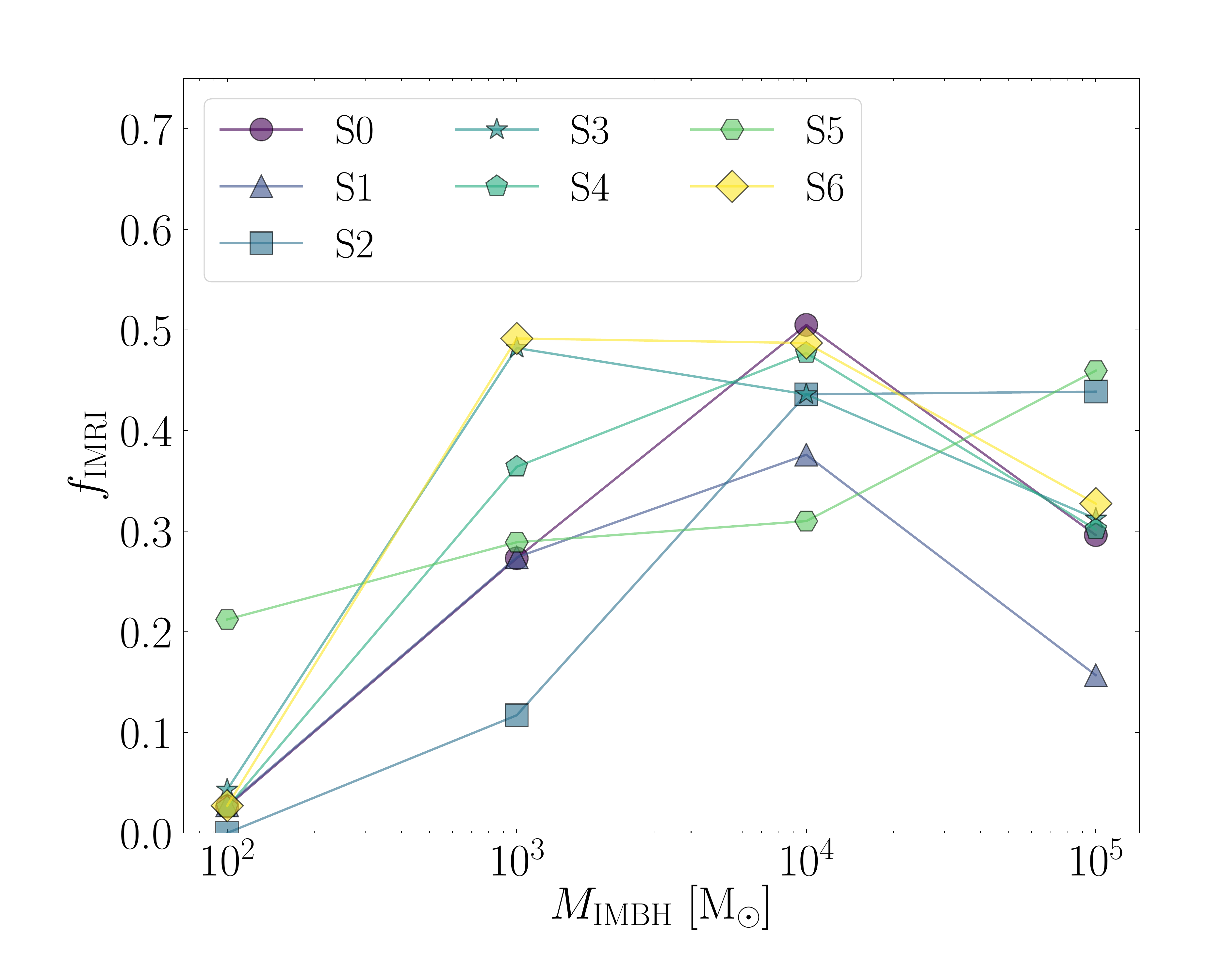}
    \caption{IMRIs merger probability as a function of the IMBH mass for all models explored.}
	\label{fig:gwprob}
\end{figure}

As summarised in Table \ref{tab:mergers}, the fraction of systems resulting in a IMRI depends on the IMBH mass and the model adopted. Figure \ref{fig:gwprob} shows the fraction of mergers $f_{\imri}$ as a function of the IMBH mass for different models. We see that the scatter among different models and for a fixed IMBH mass value is considerable, spanning from $f_{\imri} \sim 10\%$ (S2) to $50 \%$ (S6) for $M_\ibh = 10^3\Ms$. 

Despite the absence of a clear trend, our results suggest that the IMRI merger probability maximizes at IMBH masses $M_\ibh\sim 10^4\Ms$, while it is limited to a few percent in the case $M_{\ibh} = 100\Ms$. This owes primarily to the fact that models with $M_\ibh = 100\Ms$ have on average wider orbits, and they can move more in the cluster potential compared to heavier IMBHs. Conversely, models with heavier IMBHs are characterised by initially tighter orbits, a deeper potential well, and the IMBH is less subjected to the Brownian motion thanks to its large inertia. This hypothesis is supported by the fact that in model S5, where the initial semimajor axis distribution is insensitive to the cluster mass, the merger fraction varies only slightly within the IMBH mass range $M_\ibh < 10^4\Ms$. Note that the majority of IMRIs formed in these models are triggered by the chaotic interactions between the three BHs, rather than by secular effects.

In the next section, we will exploit these results to infer the cosmological merger rate of IMRIs associated with different IMBH mass ranges and GW detectors.

\begin{figure*}
\centering
\includegraphics[width=0.8\textwidth]{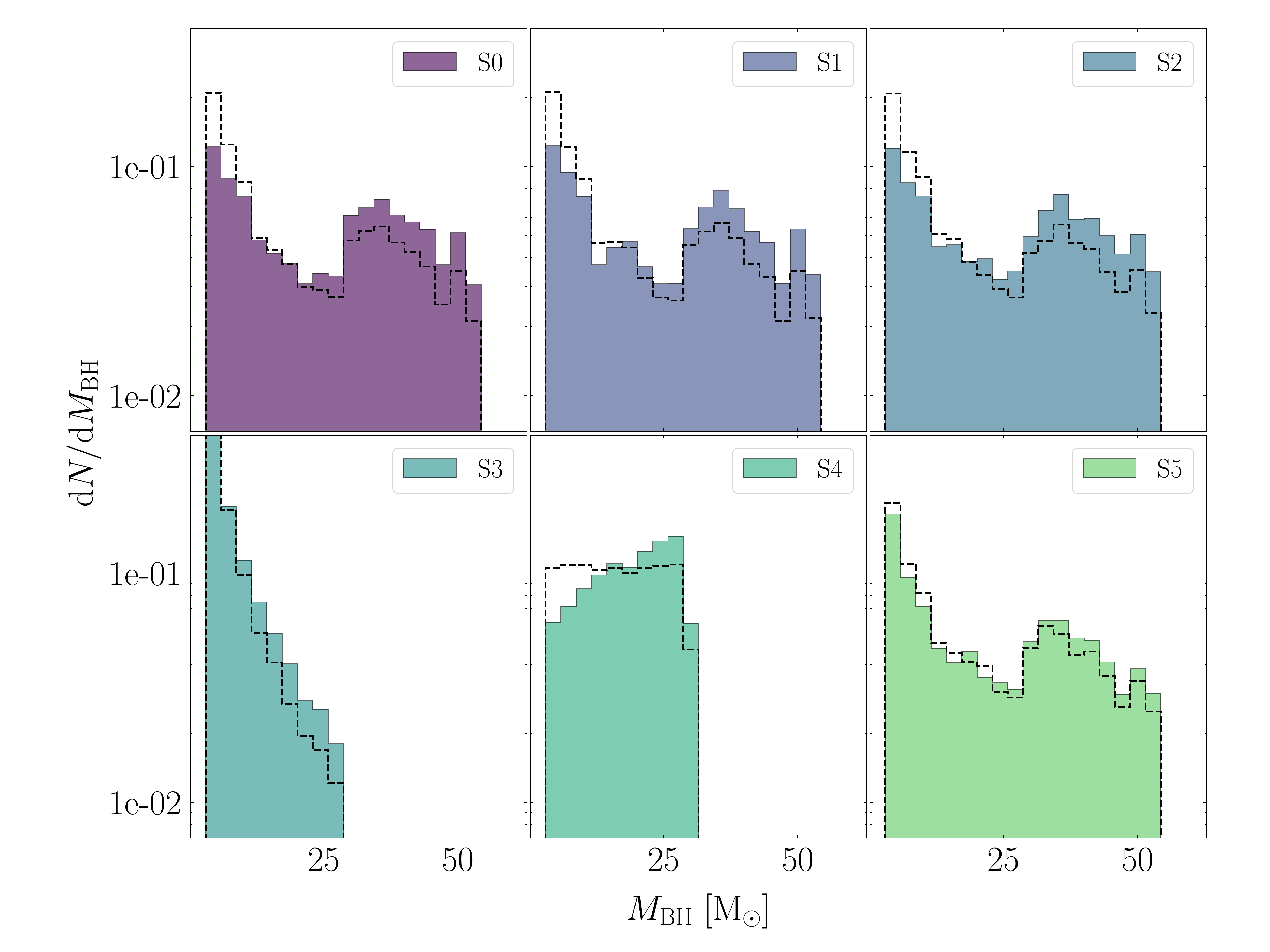}\\
    \caption{
    Mass distribution of the IMRI secondary for models S0-5. Each color corresponds to a different set to facilitate the comparison between different panels and figures. 
    Dashed lines mark the mass function adopted for stellar BHs.
    }
	\label{fig:f8}
\end{figure*}
\begin{figure}
\centering
\includegraphics[width=0.8\columnwidth]{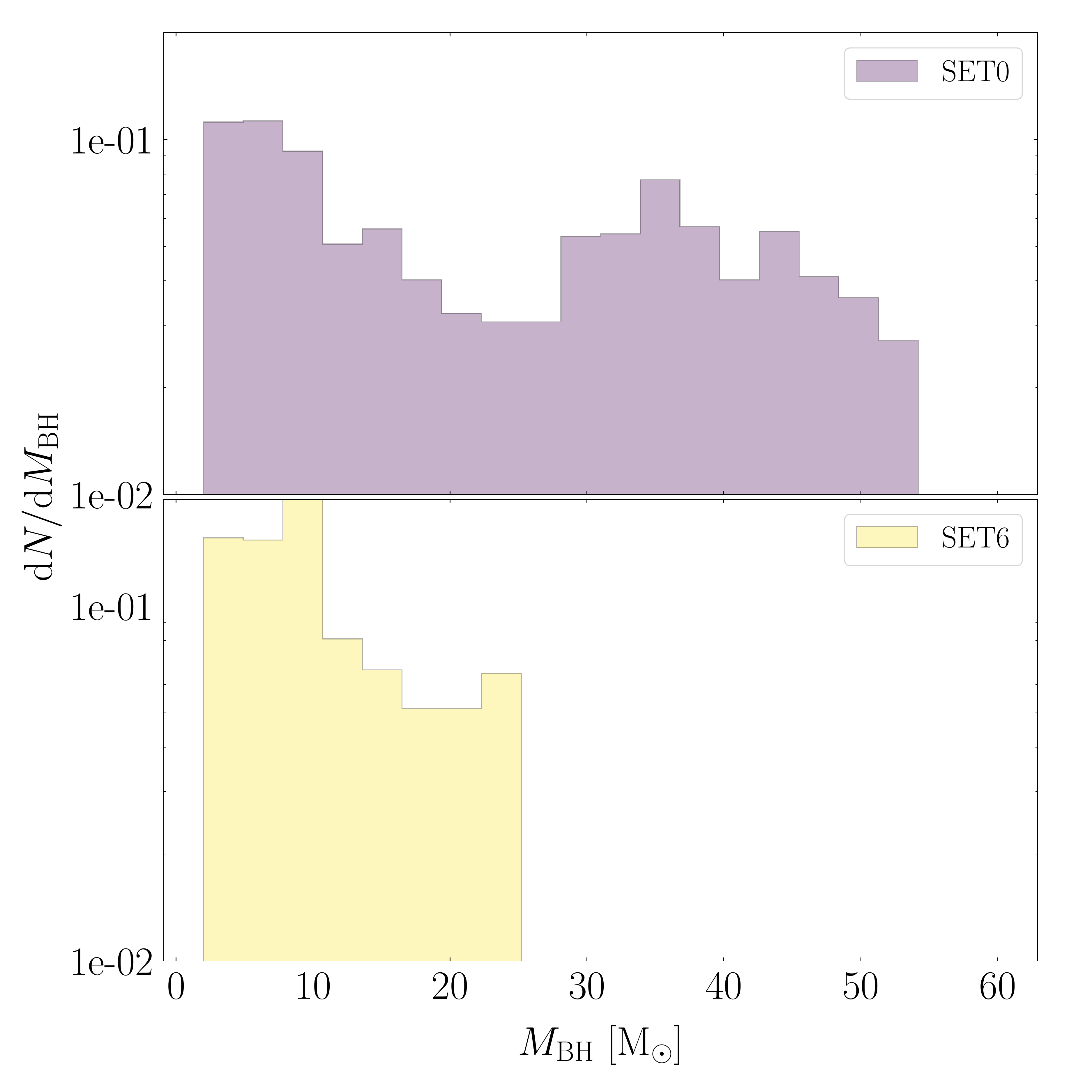}\\
\caption{
    As in Figure \ref{fig:f8}, but for dataset S0 and S6. Here we show only the mass of merging BHs.    
    }
	\label{fig:mbhdist}
\end{figure}

As shown in Figure \ref{fig:f8} for models S0-5 (different cluster density profiles and BH mass spectrum) and Figure \ref{fig:mbhdist} for models S0 and S6 (different stellar metallicity), another important property that can be inferred from our models is the mass distribution ${\rm d}N/{\rm d}M_{\bh}$ of the IMRI secondary. Comparing S0, S1, and S3 -- which differ only in the cluster density profile but adopt the same BH mass spectrum \citep{spera17} -- it is apparent that the ${\rm d}N/{\rm d}M_{\bh}$ does not depend on the environment, but rather on the BH mass spectrum adopted. In all three cases, the mass distribution shows a clearly bimodal distribution peaked at $\sim 7\Ms$ and $34\Ms$ {which is directly inherited by the assumption that BH progenitor masses are distributed according to a \cite{kroupa01} initial mass function and that the natal BH mass spectrum follows \cite{spera17}}. The picture changes significantly if another BH mass spectrum is adopted. In the case of a power-law mass spectrum in the range $M_\bh = 3-30\Ms$ we find that ${\rm d}N/{\rm d}M_{\bh}$ declines sharply starting from $3\Ms$ and truncates at $> 25 \Ms$, whereas in the case of an initially flat mass spectrum the BH mass distribution increases toward $25 \Ms$ and abruptly drops beyond this value. 

Assuming smaller values for the initial semimajor axis (S5) does not affect significantly the merging BH mass distribution, which in fact resembles that obtained for S0-3. As shown in Figure \ref{fig:mbhdist}, increasing the stellar metallicity to solar values (S6) implies a reduction of the maximum value of the IMRI secondary mass to $25\Ms$. 

Our models show that the mass distribution of IMRIs' secondary reflects the underlying BH mass spectrum, unless this is considerably flat. This suggests that detecting IMRIs can help unravel the features of BH populations lurking in dense clusters. For instance, the double peak distribution apparent in S0-2 and S5 is essentially due to the stellar evolution recipes adopted for single stars. If dynamics did not shape significantly the population of BHs around an IMBH, IMRIs can help us unravelling stellar BH natal mass spectrum. On the other hand, if dynamics had enough time to affect significantly the BH population -- e.g. causing the ejection of the most massive BHs via strong scattering -- the detection of IMRIs can tell us more about these {\it BH burning} mechanisms \citep[see e.g.][]{kremer20}.

\begin{figure}
\includegraphics[width=\columnwidth]{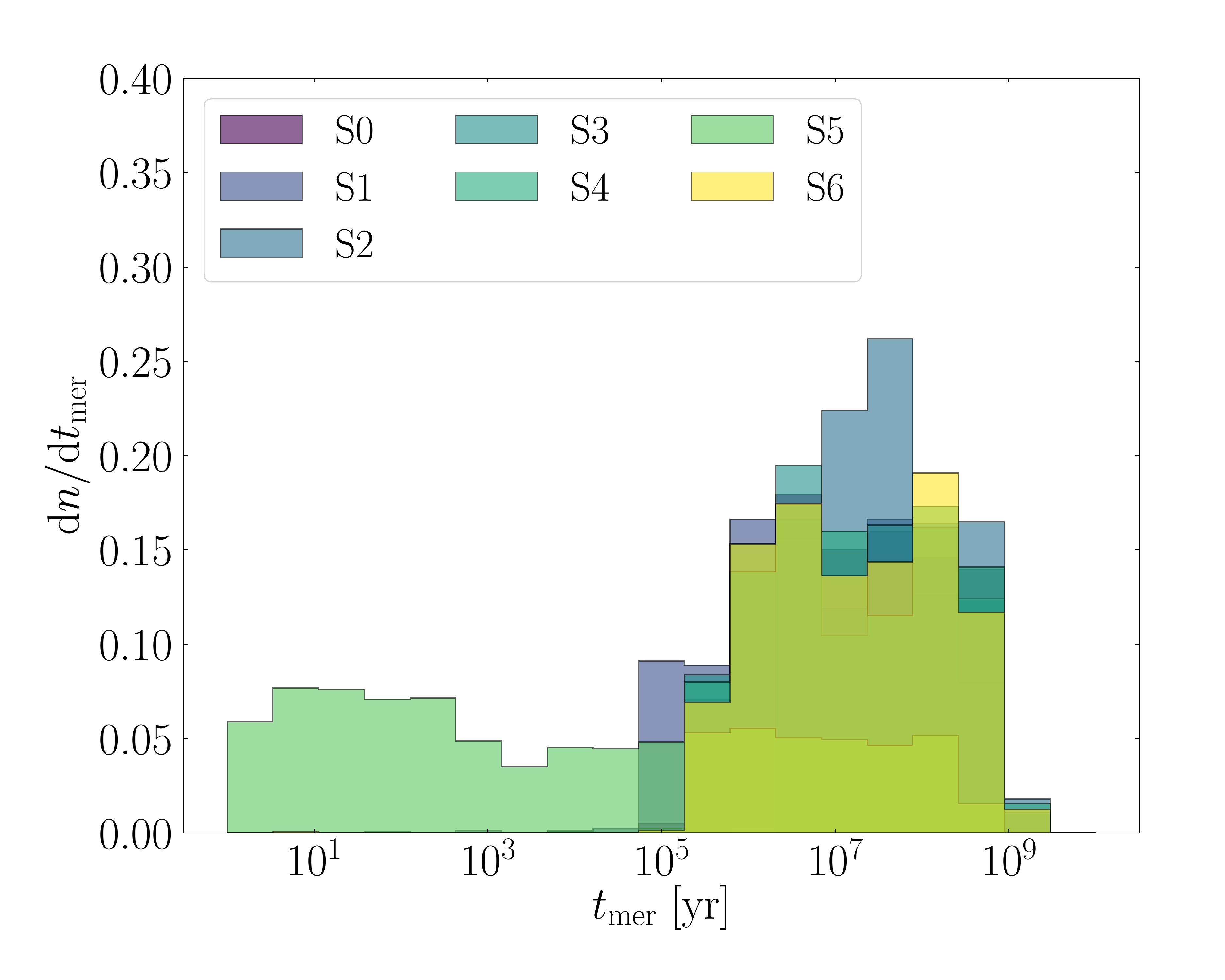}\\
\includegraphics[width=\columnwidth]{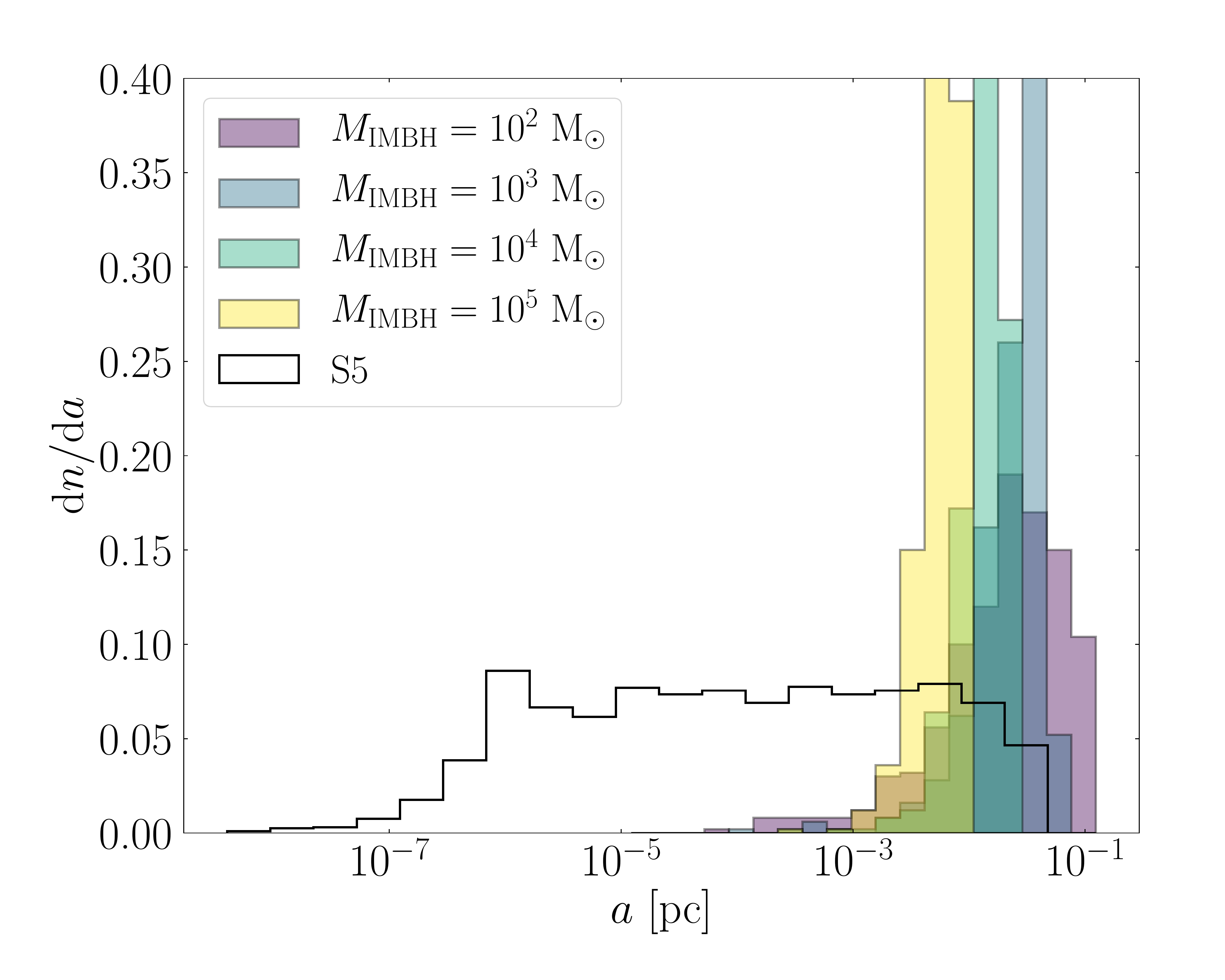}
    \caption{Top panel: distribution of IMRI merging times calculated for all mergers in all the models explored. Different colors and symbols correspond to a different model SET as indicated in the legend. Bottom panel: initial semimajor axis distribution in set S5 (black steps) and in set S0 (filled steps) differentiated through the IMBH mass.}
	\label{fig:f10a}
\end{figure}

The distribution of merger times $t_{\rm mer}$, defined as the sum of the simulated time and the GW timescale evaluated through equation \ref{peters}, shown in Figure \ref{fig:f10a} highlights a clear difference between the models in which the BH orbital parameters are selected according to the underlying cluster mass distribution (S0-4 and S6) and the one in which the semimajor axis are drawn from a logarithmically flat distribution (S5). Whilst in the former case $t_{\rm mer}$ is broadly distributed in the $10^5-10^9$ yr, with a peak in correspondence of $t_{\rm mer} \sim 8\times 10^8$ yr, in the latter the $t_{\rm mer}$ distribution is generally flat in logarithmic values and extends down to $0.5$ yr. {This apparent difference owes to the adopted distribution of initial semimajor axis, wich in S5 is logarithmically flat in the range $0.02 - 2\times 10^{4}$ AU, whilst in all the other models the boundary values depend on the cluster mass, which is directly linked to the IMBH mass. This is clearly shown in the bottom panel of Figure \ref{fig:f10a}, which compares the initial semimajor axis distribution adopted in S5 and S0 dataset. }

\subsection{Eccentricity of IMRIs}

\begin{figure}
\centering
\includegraphics[width=\columnwidth]{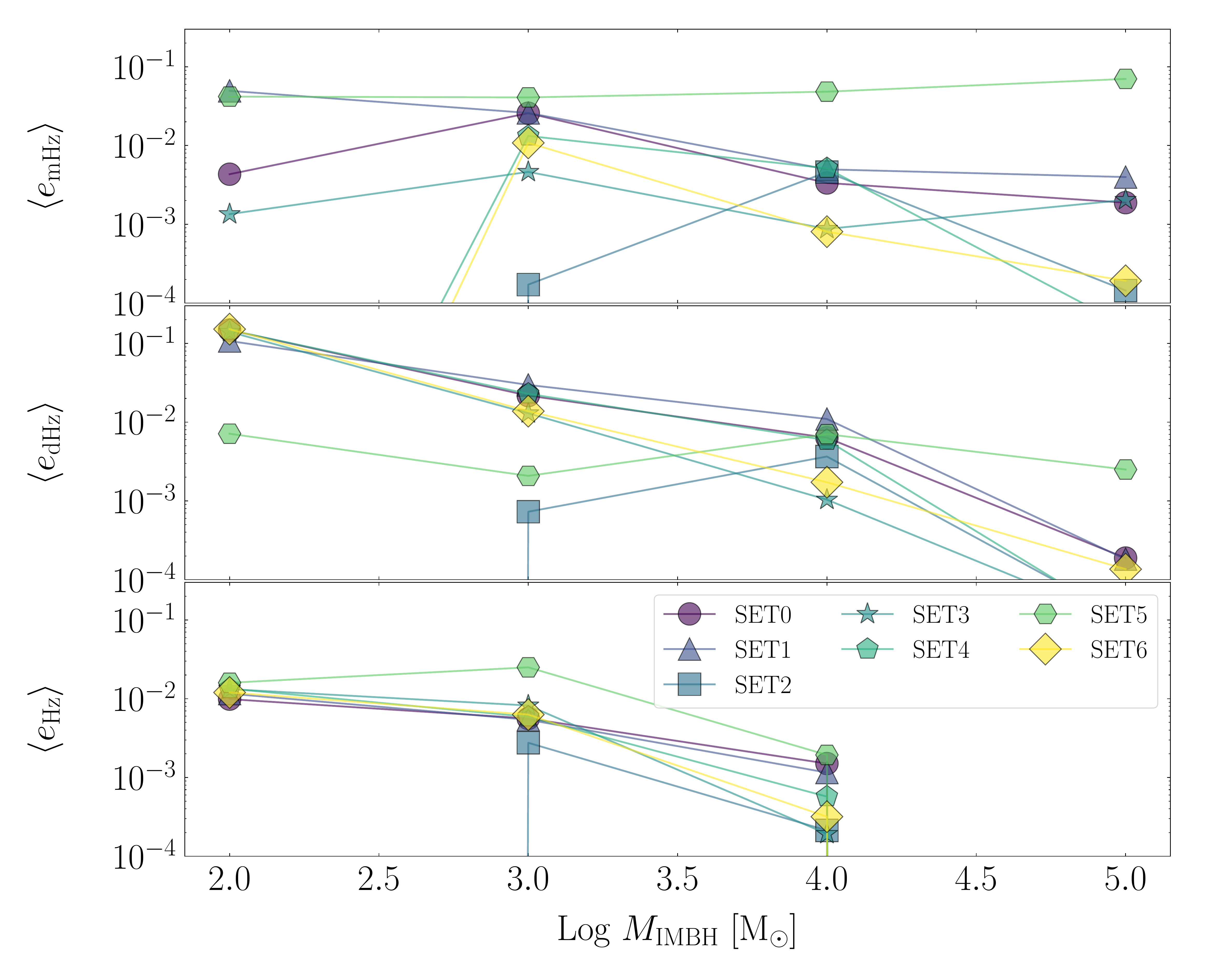}
\caption{Average eccentricity of IMRIs emitting in different frequency bands. From top to bottom panels refer to $f = 10^{-3}-10^{-1}$ Hz, $f = 10^{-1}-1$ Hz, $f = 1-10$ Hz, $f > 10$ Hz. Different colors and symbols identify different models. }
\label{fig:f10}
\end{figure}

An important parameter that could be inferred from IMRI observations is the eccentricity of the source, which could encode information about the IMRI formation channel. For instance, high eccentricities are thought to be the footprint of a dynamical origin, at least for stellar BH mergers \citep[e.g.][]{nishizawa16}.
To better understand whether our IMRI models are expected to retain a significant eccentricity while sweeping across different GW observational bands, we show in Figure \ref{fig:f10} the average value of the eccentricity calculated when IMRIs cross the frequency windows $(10^{-3}-10^{-1}-1-10)$ Hz for all simulation sets, thus covering the full range of frequencies accessible to low- (LISA), intermediate- (DECIGO), and high-frequency (LIGO, ET) detectors. It is apparent that our IMRIs have small eccentricity already at mHz frequencies, where on average we find $e=10^{-4}-0.03$.

Such low eccentricities are mainly due to the fact that the semimajor axis of the IMBH-BH merger at formation, i.e. when the effect of the third BH on the IMBH-BH evolution becomes negligible, is relatively large, i.e. $\sim 1-10$ AU. This, in turn, is connected with the initial conditions adopted. In fact, we find that in S5, where we adopt a tighter range of semimajor axis compared to all other models, the eccentricity tend to be larger at all IMBH masses and more or less at all frequencies. 

Formation channels different from the one described here tend to produce a non-negligible fraction of eccentric IMRIs, like gravitational captures or Kozai-Lidov resonances. Therefore, measuring the eccentricity of an IMRI would help unravelling its formation scenario. 

The possibility to form IMRIs that are almost circular when becoming observable to GW observatories has implications on their detectability, as we show in section \ref{gra}.

\subsection{IMBH survival in dense star clusters}
 
Promptly after a merger event, the anisotropic emission of GWs can impart a recoil kick to the merger product, depending on the mass and spin of the two merging objects \citep{campanelli07,gonzalez07,lousto08,lousto12}, which can kick out the IMBH from the cluster \citep[e.g.][]{bockelmann08,fragione17c} leading it to wander forever outside the cluster. 
This can significantly affect the retention probability of low-velocity dispersion ($5$ km/s) star clusters, which is limited to $1-5\%$ for BHs with mass $m_\bh \sim 100 \Ms$ that undergo 1-2 consecutive mergers \citep{arca20}. 

Assessing the retention probability for IMBHs is crucial to place constraints on their possible presence in star clusters. To explore this aspect, we perform a statistical analysis on our models to determine the retention probability of IMBHs that undergo an IMRI phase in star clusters.

For each merger in any modelled set, we use the numerical relativity fitting formulae provided by \cite{jimenez17} to calculate the remnant IMBH mass, spin, and effective spin parameter, and adopt \cite{lousto12} prescriptions to calculate the GW recoil kick, following the implementation described in \cite{arca20}. 

Since the kick is intrinsically linked to the component spins direction and amplitude, we proceed as follows. We assign to the IMBH an initial spin either $S_\ibh = 0.01$ or $S_\ibh = 0.99$, so to explore the regime of an almost non-rotating or maximal IMBH, whereas for the stellar BH natal spin distribution we assume a Gaussian centred on $S_{\bh} = 0.5$ with a dispersion of $0.1$\footnote{We found that a different choice for the stellar BH natal spin distribution does not affect sensibly the results.}. 

The recoil velocity is compared with the escape velocity calculated in the centre of the host cluster, which can be derived from the adopted cluster potential, i.e. $v_{\rm esc}^2 = 2\phi(0)$, a quantity that can be connected to the cluster mass $M_\gc$, the half-mass radius $R_h$, and the inner slope of the density profile $\gamma$ through simple formula, e.g.:
\begin{equation}
v_{\rm esc}^2 = 
	\begin{cases}
		\displaystyle{\frac{G M_\gc}{R_h (2-\gamma)[2^{1/(3-\gamma)} - 1]}}, & {\rm ~Dehnen~(1993),}  \vspace{0.2cm}\\
		\displaystyle{\frac{1.3 G M_\gc }{R_h}},                                & {\rm ~Plummer~(1915).}
	\end{cases}	
\label{eqesc}
\end{equation}

We find that the IMBH retention is ensured whenever its mass exceeds $M_\ibh \geq 10^4\Ms$, due to the fact that the host cluster escape velocity is expected in the range $80-200$ km s$^{-1}$ and the recoil kick is generally limited to $< 1$ km s$^{-1}$ due to the IMRI small mass ratio. However, the picture is more complicated for more modest IMBH masses.

\begin{figure}
\centering
\includegraphics[width=\columnwidth]{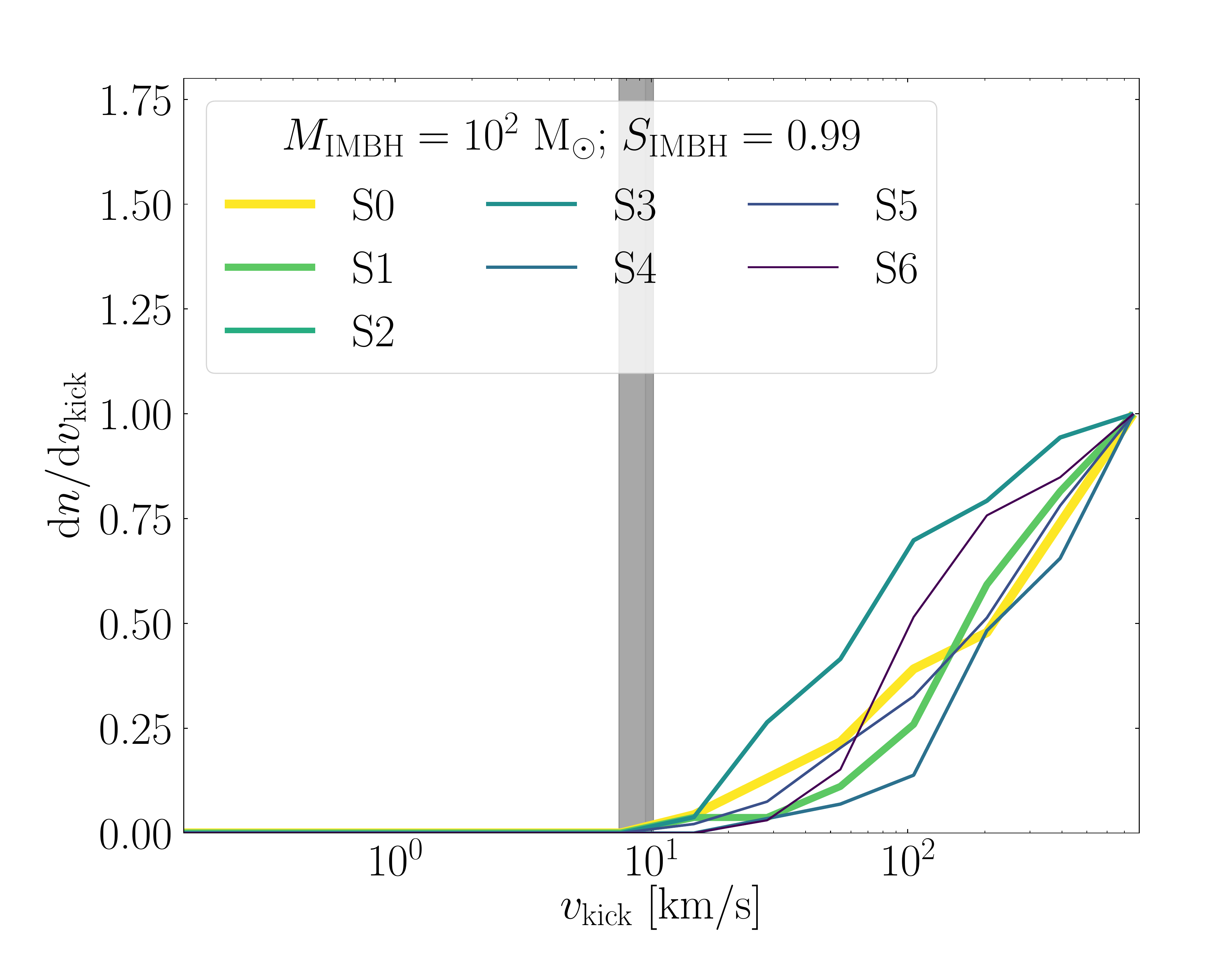}\\
\includegraphics[width=\columnwidth]{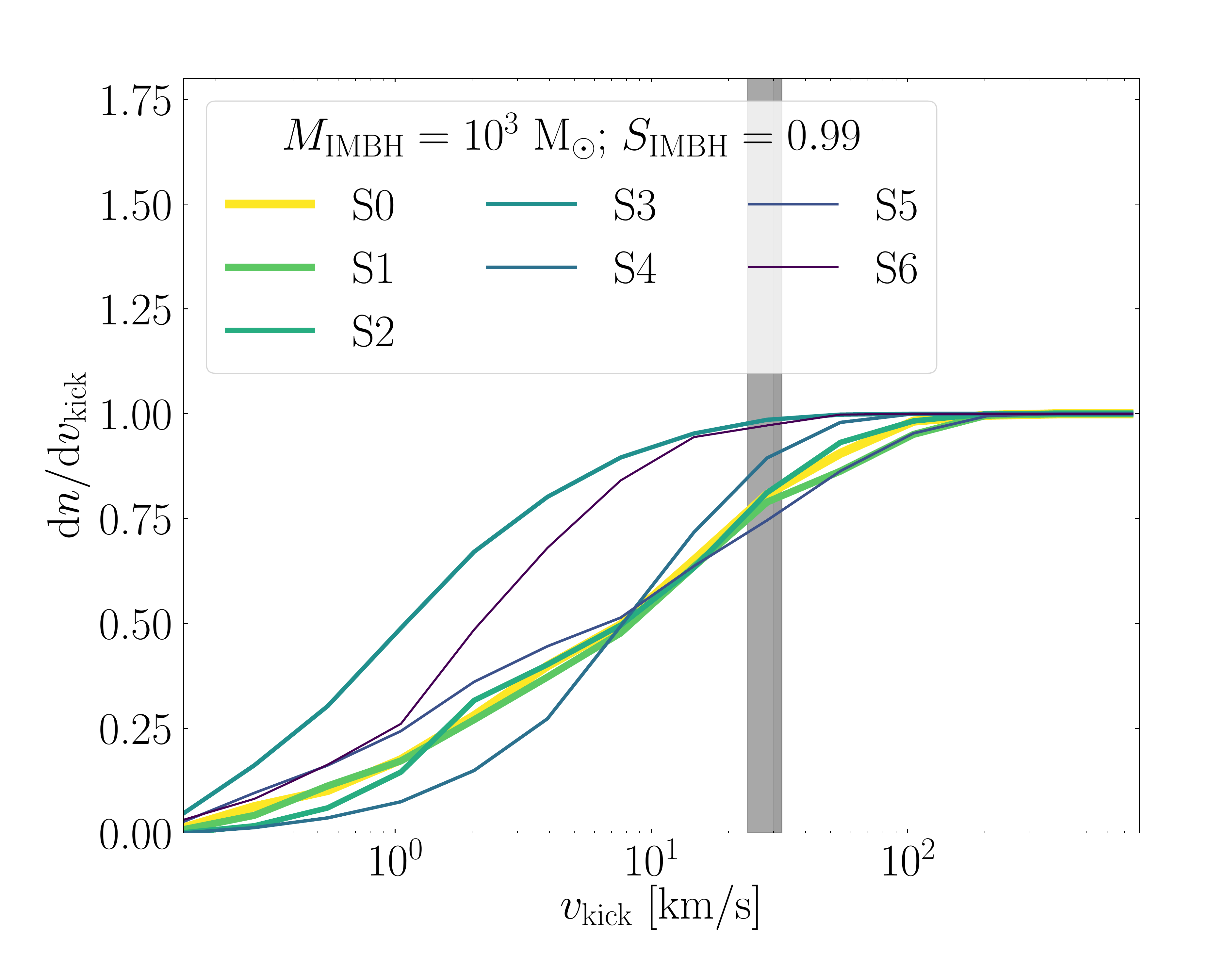}
\caption{Cumulative distribution of GW recoil kicks assuming an IMBH spin $S_\ibh = 0.99$ and assuming in top(bottom) panel an IMBH mass $M_\ibh = 10^2(10^3)\Ms$. Different colors correspond to different model sets, the shaded grey area encompass the cluster central escape velocity.}
\label{fig:survival}
\end{figure}

Figure \ref{fig:survival} shows the cumulative distribution of recoil kicks for all mergers with $M_{\ibh} = (10^2-10^3)\Ms$ and assuming an IMBH initial spin $S_\ibh = 0.99$. Note that the retention probability can be directly evaluated from the cumulative distribution through the fraction of objects having $v_{\rm kick} < v_{\rm esc}$.
 
For low-mass IMBHs, $M_\ibh \sim 100\Ms$, the retention probability remains below $P_{\rm ret} = 0.5\%$ if $S_\ibh = 0.99$, and increases only slightly ($P_{\rm ret}\lesssim 1\%$) if the IMBH is slowly rotating ($S_\ibh = 0.01$). 
Note that the retention probability weakly depends on the adopted cluster structure and stellar BH mass spectrum, thus suggesting that the retention of IMBH remnants with masses $M_\ibh \sim 10^2\Ms$ in clusters with a mass $M_\gc \sim 10^4\Ms$ is highly unlikely unless the host cluster central velocity dispersion exceeds $30-50\kms$.

Note that assuming that the IMBH with $M_\ibh = 10^2\Ms$ form in a cluster ten times heavier than the adopted value (i.e. $M \sim 10^5\Ms$) has little impact on the retention probability. Indeed, adopting the range of escape velocities calculated for $M = 1.7\times 10^5\Ms$ clusters and assuming $M_\ibh = 100\Ms$ leads to a retention probability $P_{\rm ret} \sim 10\%$ in all models set but S3, for which $P_{\rm ret} \lesssim 25\%$.    

At IMBH masses $M_\ibh \sim 10^3\Ms$, instead, we see that the retention probability ranges between $P_{\rm ret} = 75-99\%$ depending on the adopted BH mass spectrum, being $P_{\rm ret}$ maximized in the case of S3, i.e. a powerlaw mass function, and S6, i.e. for BHs with solar metallicity progenitors. In both cases, BH masses are on average lower than in other models, thus they will lead to IMRIs with smaller mass ratios that, consequently, receive smaller kicks.
Our models suggest that in a population of metal-poor clusters with masses typical of globular clusters (i.e. $> 10^5\Ms$) there is more than $75\%$ probability to retain an IMRI remnant. 

\begin{figure}
\centering
\includegraphics[width=\columnwidth]{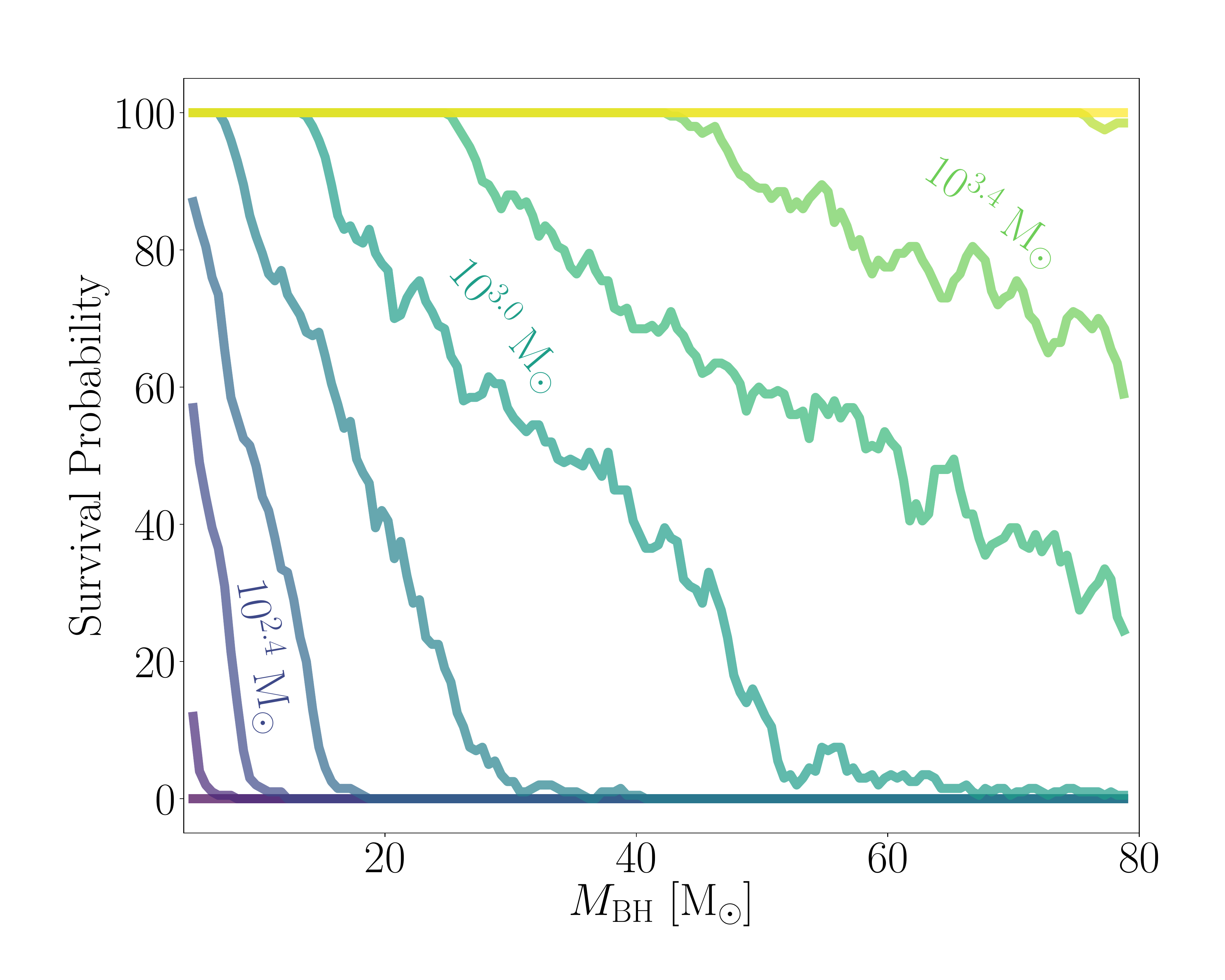}
\caption{IMBH retention probability as a function of the merging BH mass for different values of the IMBH mass identified by the color-coding.}
\label{fig:survival2}
\end{figure}

To further highlight the role of the stellar BH in determining the retention of the IMRI remnant, we proceeded as follows:
\begin{enumerate}
\item we divide the IMBH mass range -- $M_\ibh = 10^2-5\times 10^5\Ms$ -- in 15 values evenly distributed in logarithmic values;
\item for each IMBH mass, we create a sample of 100 stellar BHs whose masses are calculated through \cite{spera17}, assuming for the BH progenitors a \cite{kroupa01} initial mass function and a metallicity of $Z = 0.0002$, i.e. the same adopted in models S0-2;
\item we assume an IMBH spin $S_\ibh = 0.99$, whilst for the BH we extract the spin from a Gaussian centred on $S_\bh = 0.5$ with dispersion $0.1$;
\item we use \cite{jimenez17} numerical relativity fitting formulae to calculate the IMRI merger remnant mass $M_{\rm rem}$, spin $S_{\rm rem}$, and recoil velocity $v_{\rm kick}$, following the procedure depicted in \cite{arca20}.
\end{enumerate}
For each $M_\ibh -m_\bh$ pair, we repeat 100 times steps 3 and 4 and, for each of them, we check the $v_{\rm kick} < v_{\rm esc}$ conditions, with $v_{\rm esc}$ ranging between $10-250\kms$ for the IMBH mass range $M_\ibh = 10^2-10^5\Ms$. The retention probability obtained through the procedure above is shown in Figure \ref{fig:survival2}.

We find that an IMBH with mass $M_\ibh < 200 \Ms$ gets ejected whenever the merging companion has a mass $m_\bh \geq 10 \Ms$. Even for heavier IMBHs, e.g. $M_\ibh = 1000\Ms$, and thus heavier clusters, the retention probability rapidly drops below $50\%$  if the companion mass exceeds $M_\bh > 40\Ms$. The retention probability attains values $>80\%$ regardless the secondary BH mass only for quite heavier IMBHs ($M_\ibh \sim 10^4\Ms$).

The analysis above can be used to place constraints on the processes that might regulate IMBH seeding and growth.
For instance, a scenario in which an IMBH forms through the merger of stellar-mass BHs seems highly unlikely in ``normal'' star clusters, given the low retention fraction of IMBHs with masses $M_\ibh < 10^3\Ms$. However, if a substantial fraction of the IMBH is assembled via stellar accretion \citep[e.g.][]{zwart02,giersz15,mapelli16,dicarlo19,rizzuto21},  IMRI formation and merger would not represent a threat to the IMBH retention and further growth.

In this sense, localizing IMRIs with LISA and similar detectors could provide us with insights on the IMBH formation history and, more in general, on IMBH formation mechanisms.

\section{IMBH evolution in dense nuclear clusters}

\subsection{IMBH spin evolution}

In this section we investigate whether the effects of a merging event can be encoded in the remnant IMBH spin, and whether it is possible to use this quantity to infer IMBH evolutionary pathways.

For each merger in all our models, we assign to the BH a spin drawn from a Gaussian distribution centred at $S_\bh=0.5$ with a dispersion of $0.1$, and we assume either an almost non-rotating ($S_\ibh = 0.01$) or nearly extremal ($S_\ibh=0.99$) IMBH. 
 
\begin{figure}
\centering
\includegraphics[width=\columnwidth]{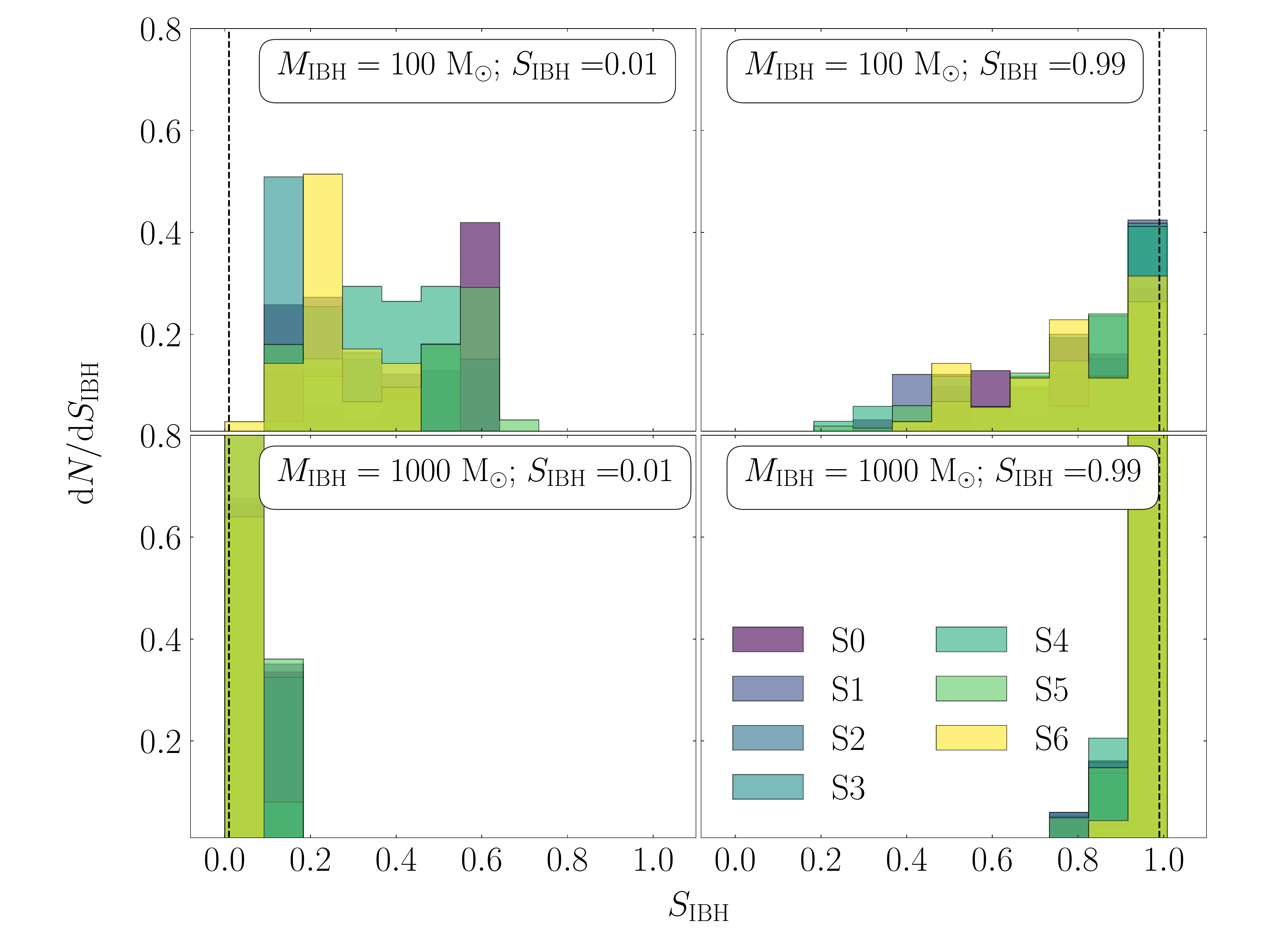}
\caption{Spin distribution of IMBH remnants in our simulations for all models and for an IMBH mass of $10^2\Ms$ (top row) and $10^3\Ms$ (bottom row), assuming that the initial IMBH mass is either $0.01$ (left column) or $0.99$ (right column).}
\label{fig:spin1}
\end{figure}

We find that a single merger event does not affect sensibly the IMBH spin for mass values $M_\ibh > 10^4\Ms$, but it leaves a clear imprint in the spin of lighter IMBHs. Figure \ref{fig:spin1} shows the remnant spin for mergers in simulations assuming an IMBH mass of $M_\ibh = (10^2-10^3)\Ms$ and spin $S_\ibh = (0.01-0.99)$. The panels make clear that even a single merger event can change significantly the remnant IMBH spin. 

In the case of low-mass, slowly spinning IMBHs ($M_\ibh,S_\ibh = 10^2\Ms,~0.01$) the remnant IMBH spin can attain values as large as $S_{\rm rem} = 0.8$, while for rotating IMBHs the remnant spin shows a steep rise from 0.2 to 1. 
The difference is even more apparent in the case of $M_\ibh = 10^3\Ms$, since in this case the spin can increase to up to $S_{\rm rem} = 0.2$ if the IMBH is non-rotating, or reduce down to $S_{\rm rem} = 0.8$ if its initial spin is close to unity. This clear difference could unravel the history of the IMBH, in particular if the GW kick is sufficiently large to eject the remnant from the cluster and thus prevent it to undergo further merger events.

\begin{figure}
\centering
\includegraphics[width=\columnwidth]{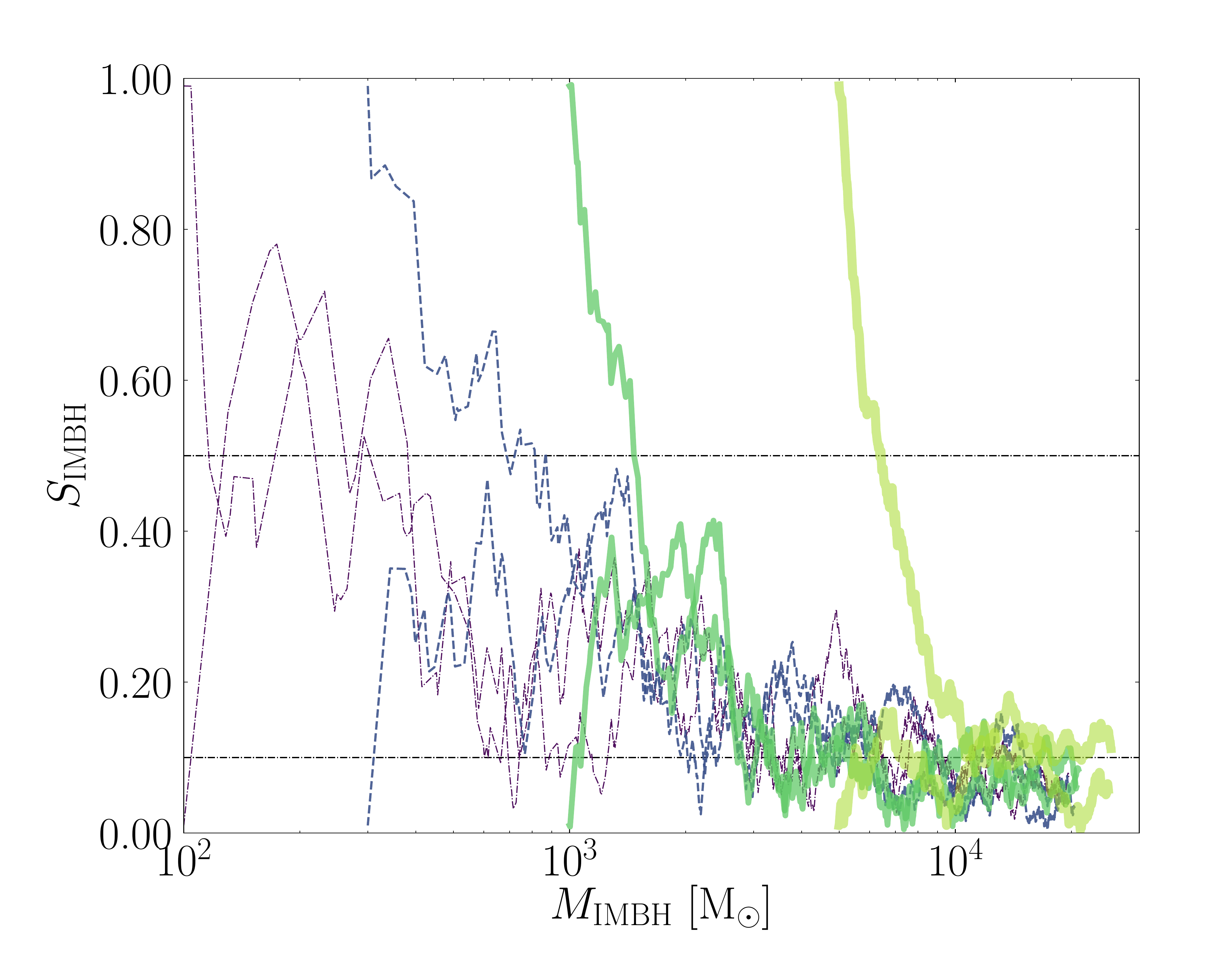}
\caption{Spin evolution for IMBH seeds that undergo a series of merger with stellar BHs. The increase in mass marks the direction of time. The leftmost point in each curve represents the adopted values of IMBH's initial mass and spin. }
\label{fig:spin2}
\end{figure}

However, it is worth exploring how the IMBH spin would change if it was initially formed in a cluster sufficiently large to retain the remnant after every merger. Figure \ref{fig:spin2} shows the spin variation of an IMBH with initial mass either $M_\ibh = 100-300-1,000-5,000\Ms$ and initial spin either $S_\ibh = 0.01-0.99$ that undergoes multiple mergers. The companion BH progenitor mass is extracted from a Kroupa IMF and the BH mass is taken from \cite{spera17}, whereas its spin is drawn by a Gaussian centred in $S_\bh = 0.5$ with dispersion $0.1$. The plot makes clear that, regardless of the IMBH initial mass and spin, after a certain number of merging events the remnant spin tends to attain a value $S_{\rm rem}  = 0-0.2$.  

Such a result could have implications on the IMBH formation history. Let's assume a simple toy model in which an IMBH seed forms from stellar evolution, thus initially $M_{\ibh} \sim 100\Ms$, and grows via multiple mergers with stellar BHs (the so-called {\it hierarchical} merger scenario). Assuming that the IMBH terminal mass is $M_{\rm rem} = 10^3\Ms$, which requires $\sim 45$ subsequent merger events, we would expect from Figure \ref{fig:spin2} a final spin $S_{\rm rem} = 0.1-0.3$. However, if the formation of a $10^3\Ms$ IMBH is driven by the direct collapse of a very massive star \citep{zwart02,giersz15,mapelli16} or the accretion of the very massive star onto a stellar BH \citep[e.g.][]{rizzuto21} the IMBH spin will be inevitably given by the process that determined either the collapse of the VMS or its accretion onto a stellar BH companion.

\subsection{IMBH seeding and growth}

\begin{figure*}
\centering
\includegraphics[width=18cm]{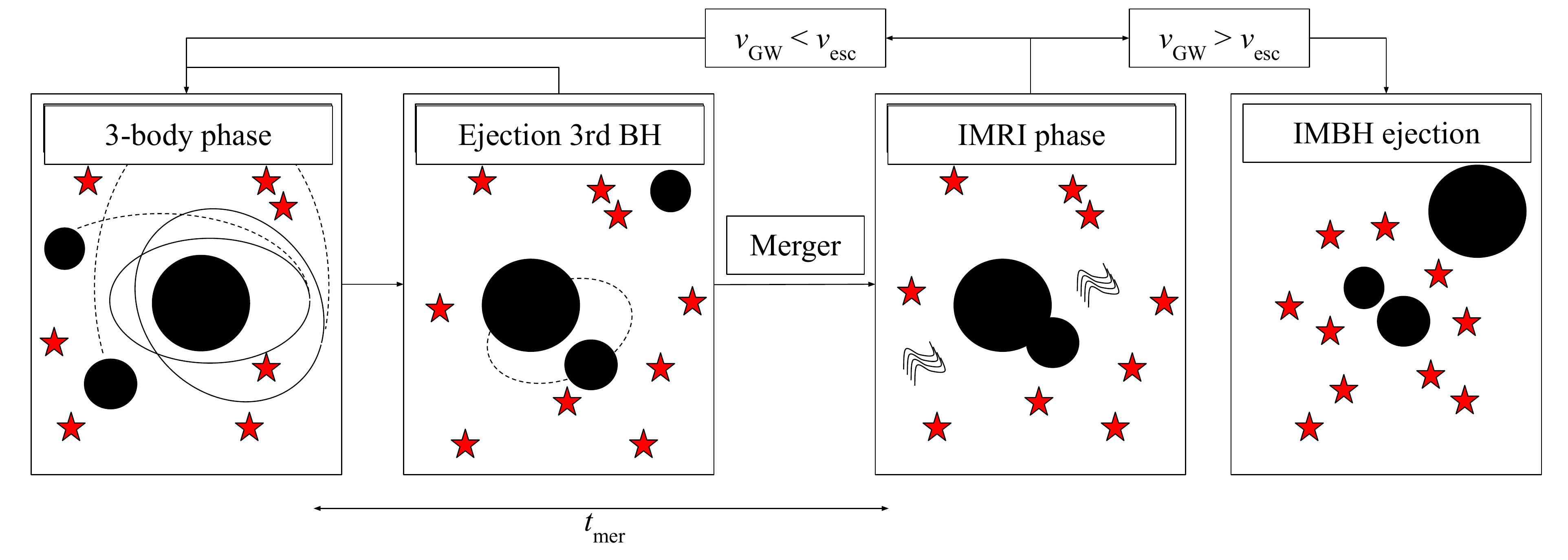}
\caption{Schematics of IMBH growth via 3-body driven IMBH-BH mergers. We assume that the 3-body phase lasts a $t_{\rm mer}$ time, after which the third BH is ejected and either the IMBH-BH binary remains stable or it undergoes an IMRI phase and ultimately merge, possibly kicking out the IMBH remnant owing to GW recoil.}
\label{scheme}
\end{figure*}

In the previous sections we have discussed how two stellar BHs can mediate the formation of an IMRI defined by an IMBH-BH binary, and how such mechanism can affect the retention and the spin evolution of the remnant IMBH. In this section, we use these results to check whether the same mechanism could support the growth of an IMBH in a dense globular or nuclear cluster, following the procedure schematized in Figure \ref{scheme} and described below. 

In order to reach this aim, we focus on an IMBH seed with mass $M_{\rm seed}$ harbored in the centre of a star cluster with central escape velocity $v_{\rm esc}$ evaluated through Equation \ref{eqesc}.

For our purposes, we adopt $v_{\rm esc} = 100$ km s$^{-1}$. Note that this implies $M_\gc > 3\times 10^6\Ms$ and $R_h < 3$ pc, regardless of the value of the slope of the density profile or the type of model adopted \citep[e.g.][]{Deh93, Plum}.

We assume that the cluster forms $t_0 = 1$ Gyr after the Big Bang, i.e. at a redshift $z\sim 5.5$\footnote{In agreement with the scenario in which globular clusters likely formed through redshift $z=2-6$ \citep[e.g.][]{forbes10,vandeberg13}}, and that an IMBH seed with mass $M_{\rm seed}$ and spin $S_{\rm seed}$ assemblies over a timescale $\sim 100$ Myr \citep{rizzuto21}. 

In the following we focus on S0 and S5, which share the same BH mass spectrum \citep{spera17} and metallicity ($Z=0.0002$) but rely on different assumptions on the IMBH-BH initial orbits. 
We assume a seed mass $M_{\rm seed} = (100-300-500)\Ms$, and a spin $S_{\rm seed} = 0.01-0.7-0.99$. We create 500 merger trees for each combination, allowing up to a maximum of $10^3$ interactions.

We assign the IMBH seed a companion BH with mass derived from the adopted mass spectrum and we assume that this binary undergoes perturbation from a third BH at a time $t_{\rm mer}$ extracted from the distribution derived from the adopted set (either S0 or S5). We assume that such interaction results in IMRI formation and merger on a statistical basis, assuming a merging probability of $p_{\rm mer} = 15\%(30\%)$, i.e. in agreement with the $p_{\rm mer}$ value derived from our simulations for S0(S5). Practically, we draw a number between $p_n = 0$ and $1$ and merge the IMBH and BH if $p_n < p_{\rm mer}$, otherwise we assume that the IMBH-BH binary remains bound and another BH comes in and, potentially, trigger the IMBH-BH merger. We also assume that, if the merger fails, the IMBH has a $30\%$ chance to exchange the previous BH with a new one, as it is suggested by our simulations. 

Note that the main difference between S0 and S5 is that in the latter the merger probability is larger and the merger time is widely distributed in the $10^{1-9}$ yr, owing to the adopted distribution of semimajor axis.  

In the case of a merger, we evaluate the recoil kick and remnant mass and spin after the merger.
This procedure is repeated until either $v_{\rm kick} > v_{\rm esc}$, the time exceeds $13.8$ Gyr, or the number of IMBH-BH interactions exceeds $10^3$. The latter criterion is motivated by the fact that in a typical stellar population, the fraction of stars that collapse to a BH is $\sim 0.07$ the total number of stars, thus implying $N_{\bh} \sim 10^3$ for a cluster with mass $M_\gc = 10^6\Ms$ and average stellar mass $m_* = 0.7\Ms$.

Although simple, this procedure enables us to rapidly check whether the mechanism studied in Section \ref{sec:imris} would be efficient enough to sustain the IMBH growth from stellar to intermediate scales. 

\begin{figure}
\centering
\includegraphics[width=0.8\columnwidth]{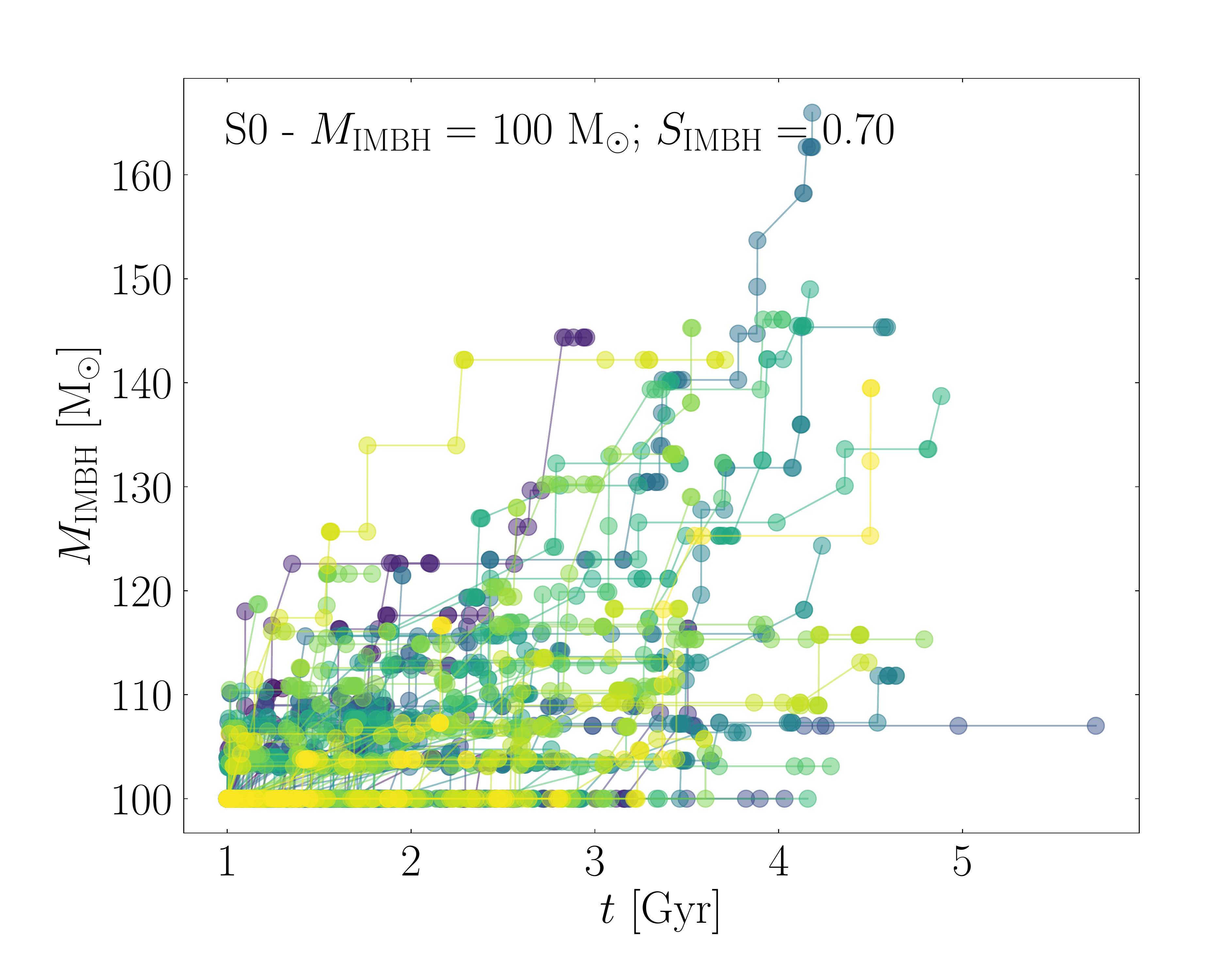}\\
\includegraphics[width=0.8\columnwidth]{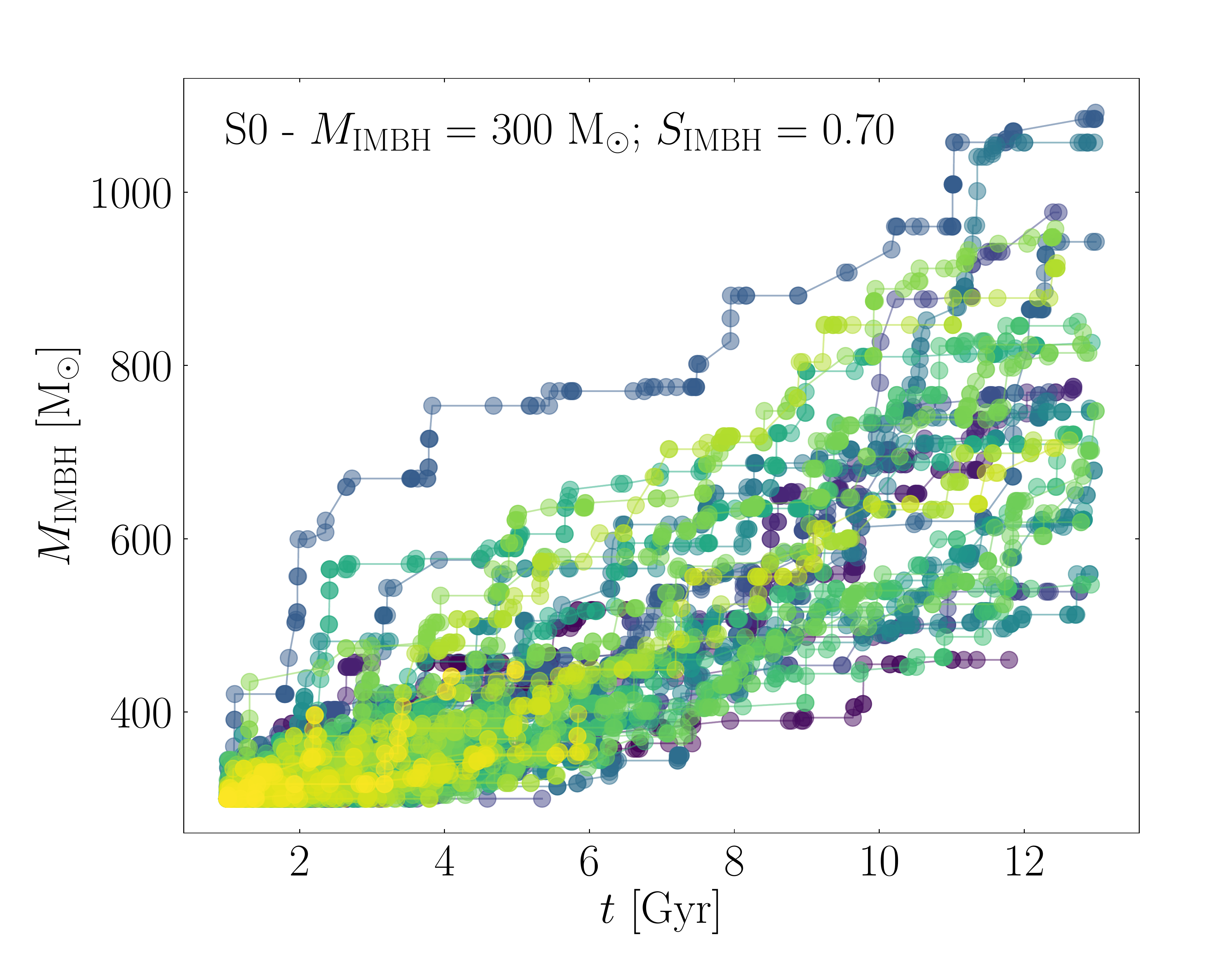}\\
\includegraphics[width=0.8\columnwidth]{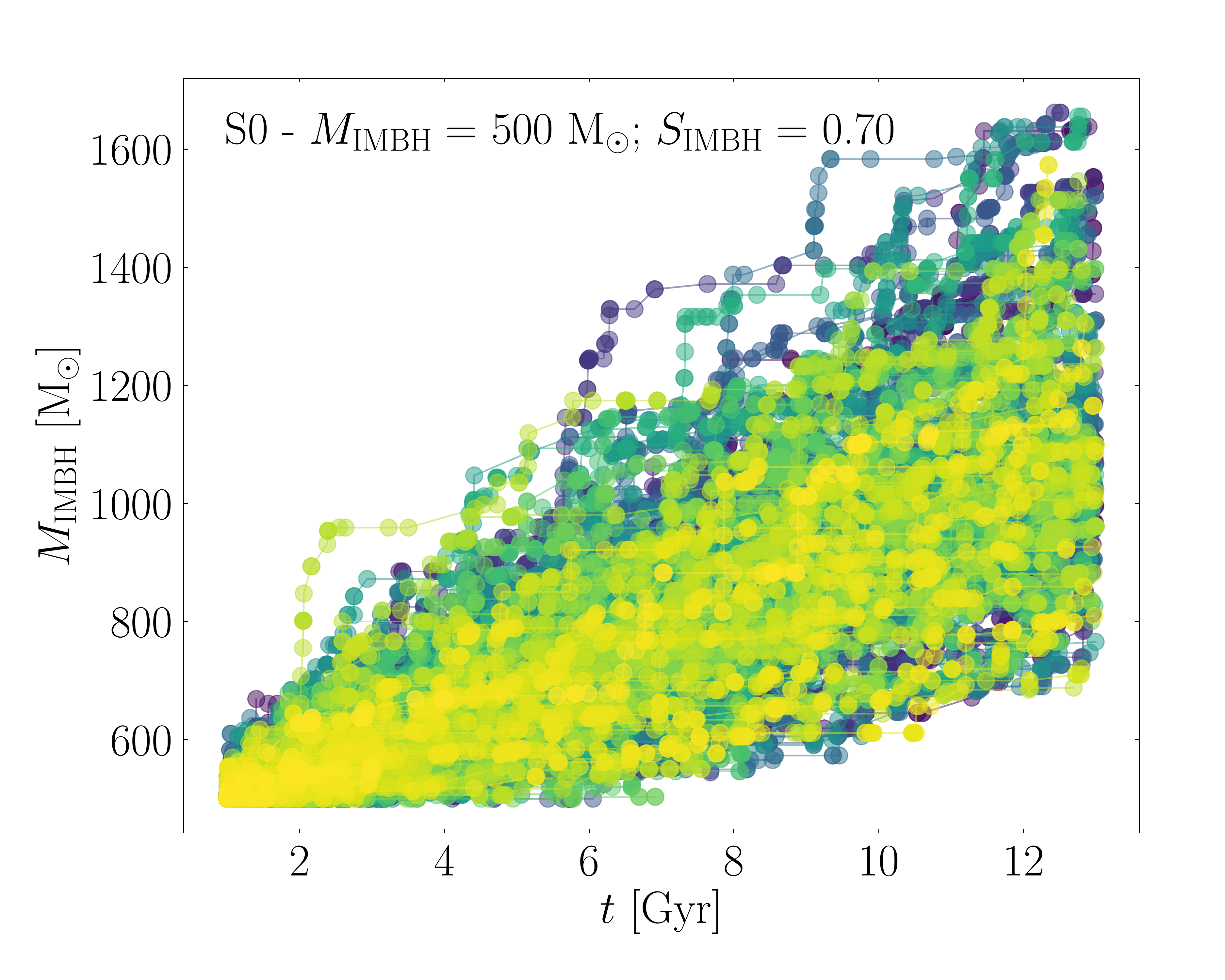}\\
\caption{Merger tree for IMBH seeds growing via 3-body dynamics in a cluster with central escape velocity $100$ km s$^{-1}$. We show only 100 tracks for each panel for readability's sake. Stellar BH masses are drawn from the mass distribution in Figure \ref{fig:f8} according to model set S0. Each track correspond to a different model. We assume that the IMBH seeding and growth starts at redshift $z=2$. }
\label{mergertree}
\end{figure}

Figure \ref{mergertree} shows a {\it merger tree} for three cases in set S0, assuming an IMBH seed spin of $S_{\rm seed} = 0.7$ and all the mass values considered. Comparing the different models highlights the importance of the IMBH seed mass in determining the IMBH growth in this particular scenario. The growth of IMBH seeds with a mass $M_{\rm seed} = 100\Ms$ is generally limited to $20-50\%$, owing to the large GW recoil kicks that tend to eject the IMBH after a few merging episodes. This limits also the timescale over which IMBHs grow, which in our models tend to be shorter than $6$ Gyr. In such a case, the three-body dynamics presented here would be inefficient in determining IMBH growth, which should thus proceed most likely via stellar accretion onto stellar mass BHs \citep[e.g.][]{giersz15}. However, if the IMBH seed initial mass is slightly larger, i.e. $M_{\rm seed} > 300-500\Ms$, the smaller GW recoil kicks enables the IMBH growth via three-body dynamics easier, leading in some cases the IMBH to reach a final mass $M_\ibh \gtrsim 10^3\Ms$ over $12$ Gyr of evolution.

\begin{figure}
\centering
\includegraphics[width=\columnwidth]{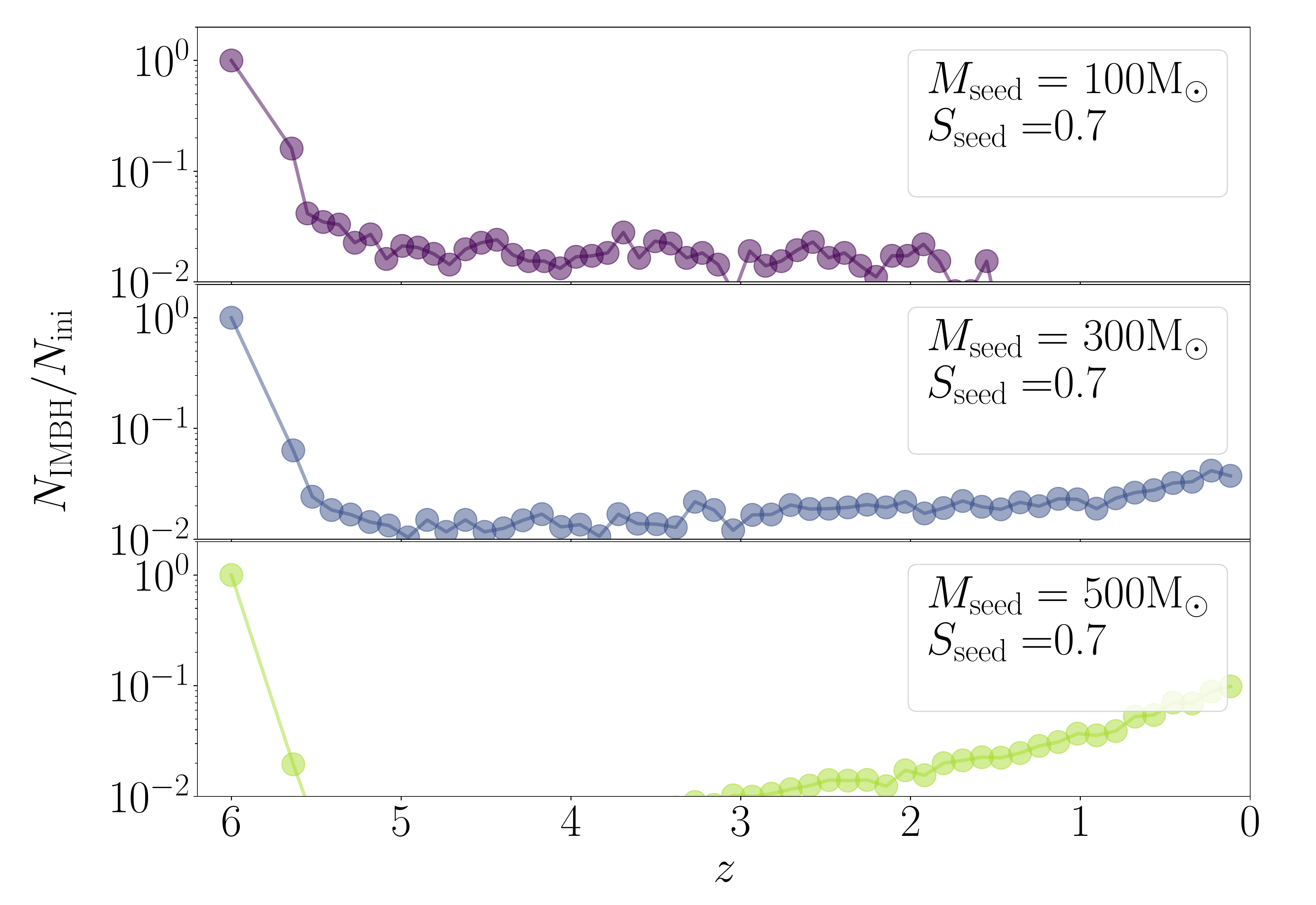}
\caption{Fraction of IMBHs at different redshift values normalized to the total number of IMBH seeds initialized at redshift $z=6$. From top to bottom, panels refer to an initial seed with mass $M_{\rm seed} = (100, ~300, ~500)\Ms$ and spin $S_{\rm seed} = 0.7$.}
\label{fig:nimbh}
\end{figure}

However, GW recoil limits the likelihood for IMBHs to grow via IMBH-BH mergers. In order to quantify the IMBH growth ``success'', we calculate the number of IMBHs that can be found at a given redshift normalized to the total number of IMBH seeds assumed. As shown in Figure \ref{fig:nimbh}, only $2-3\%$ of IMBH seeds with initial mass $M_{\rm seed} = 100\Ms$ can grow and be observed at lower redshifts, whilst this percentage raises up to $\sim 10\%$ for heavier seed ($M_{\rm seed} > 300\Ms$) and lower redshifts ($z<1$), owing to the fact that larger seeds receive smaller kicks and thus have a larger probability to undergo long merger chains. 

\begin{figure}
\centering
\includegraphics[width=\columnwidth]{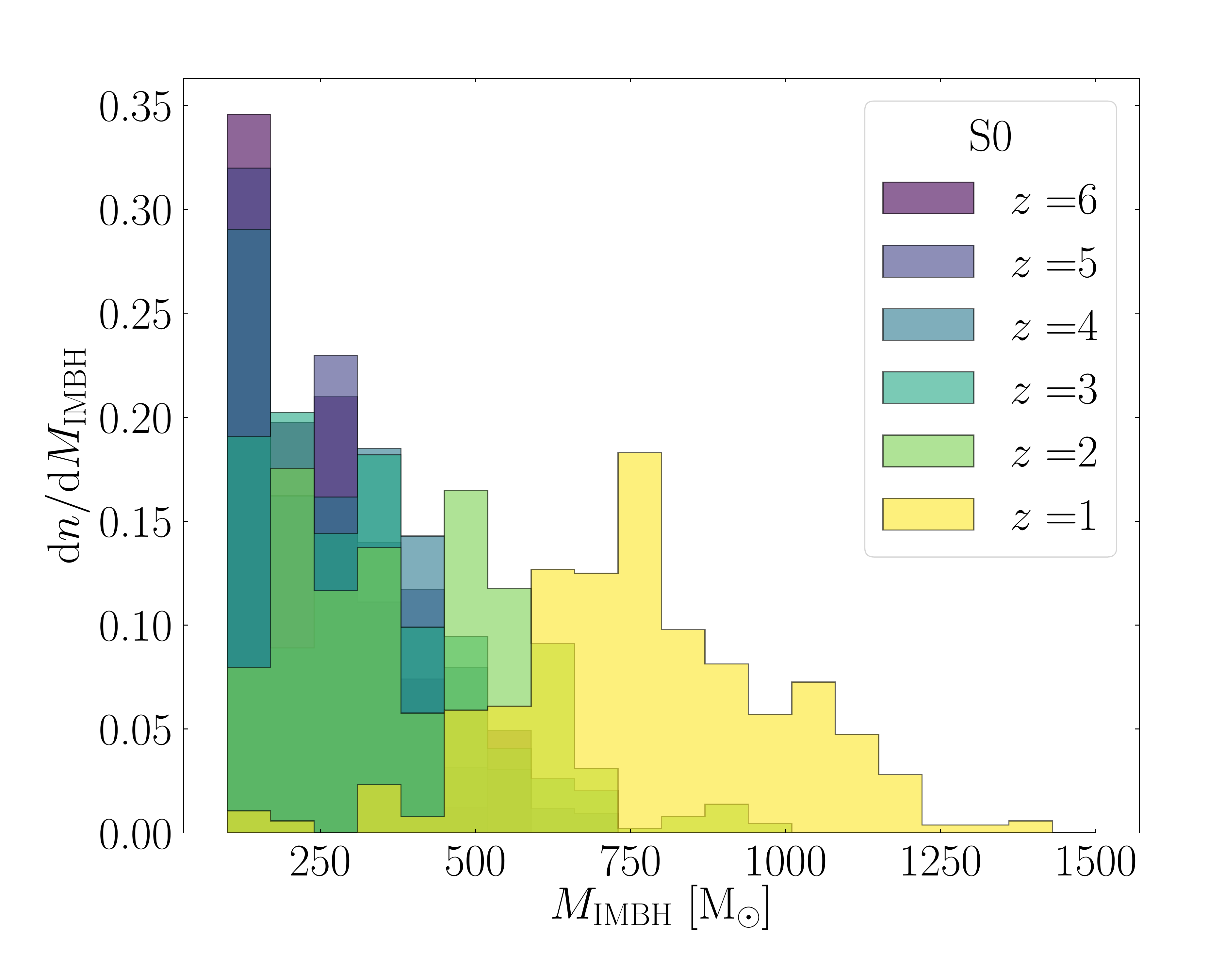}
\caption{Mass distribution of IMBHs at different redshifts, assuming that the IMBH growth process is dominated by three-body interactions. Here we consider only IMBHs that underwent at least one merger.}
\label{ibhgrw}
\end{figure}

To explore how such a mechanism would impact a putative population of IMBHs over cosmic times, we repeated the procedure above extracting the IMBH seed mass in the range $M_\ibh = (100-500)\Ms$ according to a power-law with slope $-2$ and the spin according to a Gaussian peaked in $S_\ibh = 0.5$, i.e. we assume that IMBH seeds and stellar BHs are characterised by the same spin distribution. For each model, we select the host cluster mass according to a power-law distribution with slope $-2$ in the range $M_\gc = (10^5-5\times 10^6)\Ms$ \citep{zhang99,gieles09,larsen09,chandar10a,chandar10b,chandar11}, and we assign to all clusters a half mass radius of $R = 2$ pc and central slope of the density profile $\gamma = 0.5$. Also, we assume that the host cluster formed at a redshift that is extracted randomly in the range $z= 2-6$ \citep{katz13}.
Upon these assumptions, Figure \ref{ibhgrw} shows the mass distribution of IMBHs that underwent coalescence at least once as a function of the redshift. 

We see that at redshift $z>3$ the IMBH mass distribution reflects roughly the adopted IMBH mass distribution, thus implying that mergers have a little impact on the IMBH seed masses, mostly because the number of repeated mergers at high redshifts is rather low, up to a few. At lower redshifts $z\leq 3$, instead, multiple mergers affect the IMBH mass distribution significantly, leading the present-day IMBH mass distribution to acquire a very well defined distribution peaked at $M_\ibh = 750\Ms$. Note that we are considering here only IMBHs that undergo merger with a stellar BH, thus the distribution in Figure \ref{ibhgrw} would represent the mass distribution of IMBHs as it could be seen via GW detection. 
We note the absence, in our model, of IMBHs with present-day masses larger than $1.5\times 10^3\Ms$. This is likely due to the adopted cluster mass distribution, which limit the number of heavy clusters capable of retaining the IMBH upon multiple mergers, and owing to the fact that we neglect accretion of stars as a source of IMBH growth. 

In this sense, our approach provides a simple picture of how an IMBH would grow upon multiple mergers with stellar BHs. 

On the one hand, the absence of smoking gun observations of IMBHs much heavier than $10^3\Ms$ in globular clusters would suggest that this IMBH-BH merger channel could be one of the main engine driving the (modest) IMBH growth. On the other hand, discovering the fingerprints of IMBHs with $M_\ibh \gg 10^3\Ms$ would hint at a much more complex evolutionary scenario, where either the IMBH-BH merging process proceeds more efficiently compared to our model, or a substantial fraction of the IMBH mass is accreted through stellar feeding.

In the next sections, we will exploits our model to predict the merger rate as it could be seen currently with LIGO, and in the near future with LISA, DECIGO, and the Einstein Telescope.

\section{Gravitational Waves}
\label{gra}

\subsection{Gravitational wave signal}

\begin{figure}
\centering
\includegraphics[width=\columnwidth]{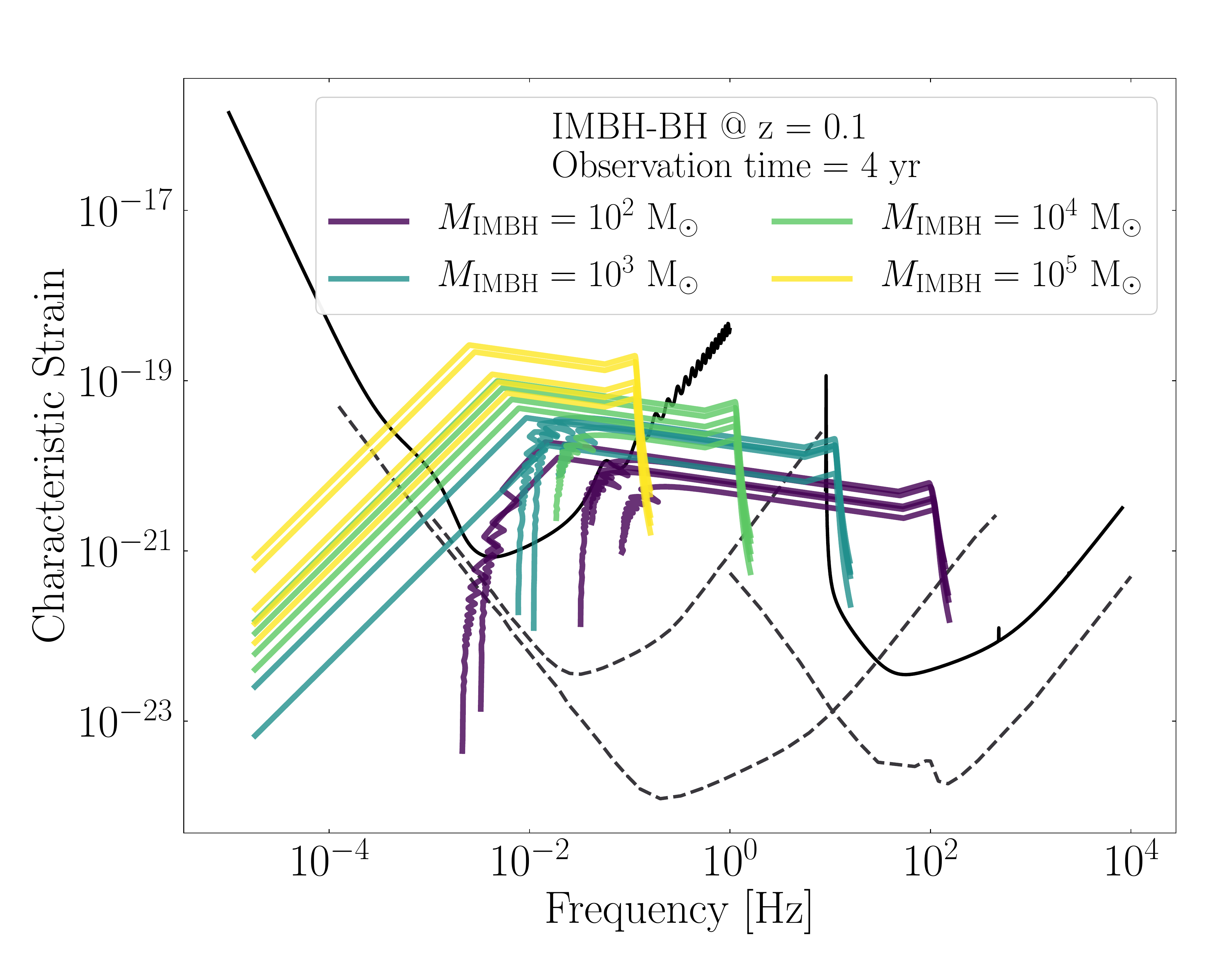}
\caption{Characteristic strain - frequency evolution for a sample of mergers in SET0. Each point in the plane refers to the signal associated with the IMRI dominant frequency. Different colors identify different IMBH mass ($M_\ibh$). The calculated signal is overlaid to the sensitivity curve of several detectors, from lower to higher frequency: LISA and LIGO (straight black line), DECIGO and Einstein Telescope, and ALIA (dashed black line).}
\label{fig:f13}
\end{figure}

Our simulations encompass a wide range of IMRI models, touching the layer of ``ordinary'' BH binaries (mass ratio $q>0.1$) and scratching the limit of extreme-mass ratio inspirals ($q < 3 \times10^{-5}$). This setup implies that the GW emission connected with our modelled mergers can populate a wide range of GW frequencies, from milli- to deci-Hz. Figure \ref{fig:f13} show the characteristic amplitude \citep{kocsis12,ASLiK18,seoane18,robson19} as a function of the frequency for a sample of IMRIs with different IMBH mass, assuming an observation time of 4 yr and a source location $z=0.1$ (i.e. at a luminosity distance $D_L\sim 460$ Mpc). We compare the modelled signal with the sensitivity curve of several detectors, namely the laser interferometer space antenna \citep[LISA,][]{eLISA13,amaro17lisa}, the advanced laser interferometer antenna \citep[ALIA,][]{bender13}, the Deciherz Gravitational-Wave Observatory \citep[DECIGO,][]{Kawamura11}, LIGO \citep{abbott16}, and the Einstein Telescope \citep[ET][]{punturo10}. We see that mergers involving IMBHs with masses $M_\ibh < 10^5\Ms$ are promising multiband sources that can be seen in one detector during the inspiral phase and in another during the merger. 

\subsection{Detecting IMBHs in Milky Way globular clusters and in the Large Magellanic Cloud}

\begin{figure}
    \centering
    \includegraphics[width=\columnwidth]{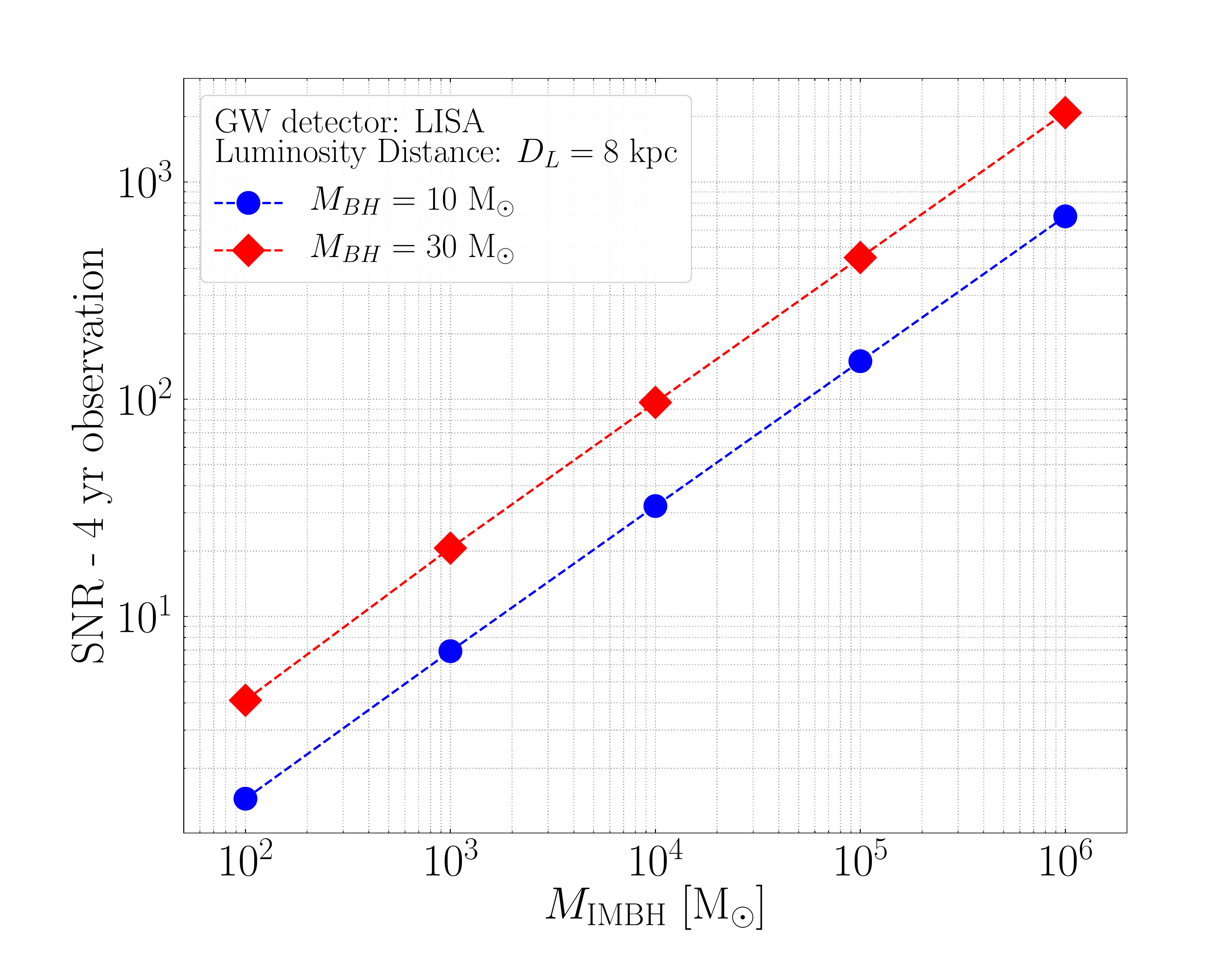}
    \caption{SNR for several IMRIs assuming an initial frequency of $1$ mHz, 4 yr of observation time, and a luminosity distance of $D_L = 8$ kpc.}
    \label{SNRMW}
\end{figure}

In this section, we exploit our results to investigate whether the current design of LISA provide enough sensitivity to unveil the presence of IMBHs in our closest neighbourhood. 
To perform such a study, we assume to have a nearly circular IMRI with an IMBH mass in the range $10^2-10^6\Ms$ and a BH companion weighing either $10$ or $30\Ms$. 
We assume that the IMRI emission frequency is 1 mHz, corresponding to an orbital semimajor axis $10^{-3}-10^{-2}$ AU. 
The merger time for IMRIs in this configuration ranges between $100$ and $10^4$ yr, thus much larger than the observation time. Assuming 4 yr observation time, the IMRI frequency will not vary sensibly, being the frequency variation $\Delta \ln f < 10^{-4}$. As noted by \cite{robson19}, whenever the latter quantity remains below 0.5, a GW source should be treated as nearly stationary. Upon this assumption, the signal-to-noise (SNR) ratio can be written as
\begin{equation}
    {\rm SNR}^2 = h_{\rm GB}^2 f / h_n,
\end{equation}
where $f$ is the source initial frequency, $h_n$ is the adimensional instrument sensitivity, and 
\begin{equation*}
    h_{\rm GB} = \frac{8T_{\rm obs}^{1/2} (G\mathcal{M}/c^3)^{5/3} (\pi f)^{2/3}}{\sqrt{5}D_L/c},
\end{equation*}
with $T_{\rm obs}$ the observation time, $\mathcal{M}$ the IMRI chirp mass, and $D_L$ the luminosity distance. Note that the relation above implies that the SNR scales with $D_L^{-1/2}$.
Figure \ref{SNRMW} shows the SNR calculated for the LISA instrument for IMRIs having an initial frequency $f=1$ mHz and masses in the $10^2-10^6 \Ms$ range, assuming a luminosity distance of $D_L = 8$ kpc. 
We find that an IMRI with mass $M_\imri = (30 + 10^3)\Ms$ has an associated SNR of 20, whereas this quantity rises up to 100 for a $10^4\Ms$ IMBH. 
These estimates suggest that LISA can play a crucial role in probing the existence of IMBHs in the Galactic backyard. 

Over the last decade a number of works pointed out the potential presence of IMBHs in Galactic globular clusters, although in most cases the results were inconclusive. The family of clusters orbiting closer than 8 kpc to us include several IMBH host candidates: 47Tuc \citep{kiziltan17}, NGC6266 \citep{abbate19}, NGC6128 and NGC288 \citep{sollima16}, NGC6388 and NGC2808 \citep{miocchi07,lanzoni07}.

However, whether these clusters actually host an IMBH is still an open question. For instance, the inferred value for the 47Tuc IMBH mass goes from $M_\ibh > 2,000\Ms$ \citep{kiziltan17} to substantially lower values $\ll 1,000\Ms$ \citep{abbate19b}, depending on the observational technique adopted. Similar arguments sustain the debates around other clusters, like NGC6388 \citep{lanzoni13,lutzgendorf13}. Therefore, LISA might play a crucial role in unravelling the presence (or not) of IMBHs in the Galaxy, at least in the closest GCs.

LISA can also enable us to probe IMBHs in nearby galaxies. For instance, it has been recently suggested that the Large Magellanic Cloud (LMC) might be harbouring an IMBH with a mass in the range $4\times 10^3-10^4\Ms$ \citep{erkal19} and up to $10^7\Ms$ \citep{boyce17}, while some of the clusters residing in the LMC can host IMBHs with masses in the range $10^3-10^4\Ms$ \citep{gualandris07}. Assuming a distance to the LMC $D_L = 49.97 \pm 1.13$ kpc \citep{pietrzy13}, we infer that LISA might observe {\it Magellanic} IMRIs with an SNR$= 8-40$, with the lower(upper) value corresponding to an IMBH with mass $10^3\Ms$($10^4\Ms$) and a BH companion of $30\Ms$.

\subsection{IMRIs merger rate}

The launch of LISA and the start of operations of the next generation of GW observatories like ET will enable us to unveil IMRIs at cosmological distances, provided that their number is sufficiently large to guarantee detection within the mission lifetime. 

\begin{figure}
\centering
\includegraphics[width=\columnwidth]{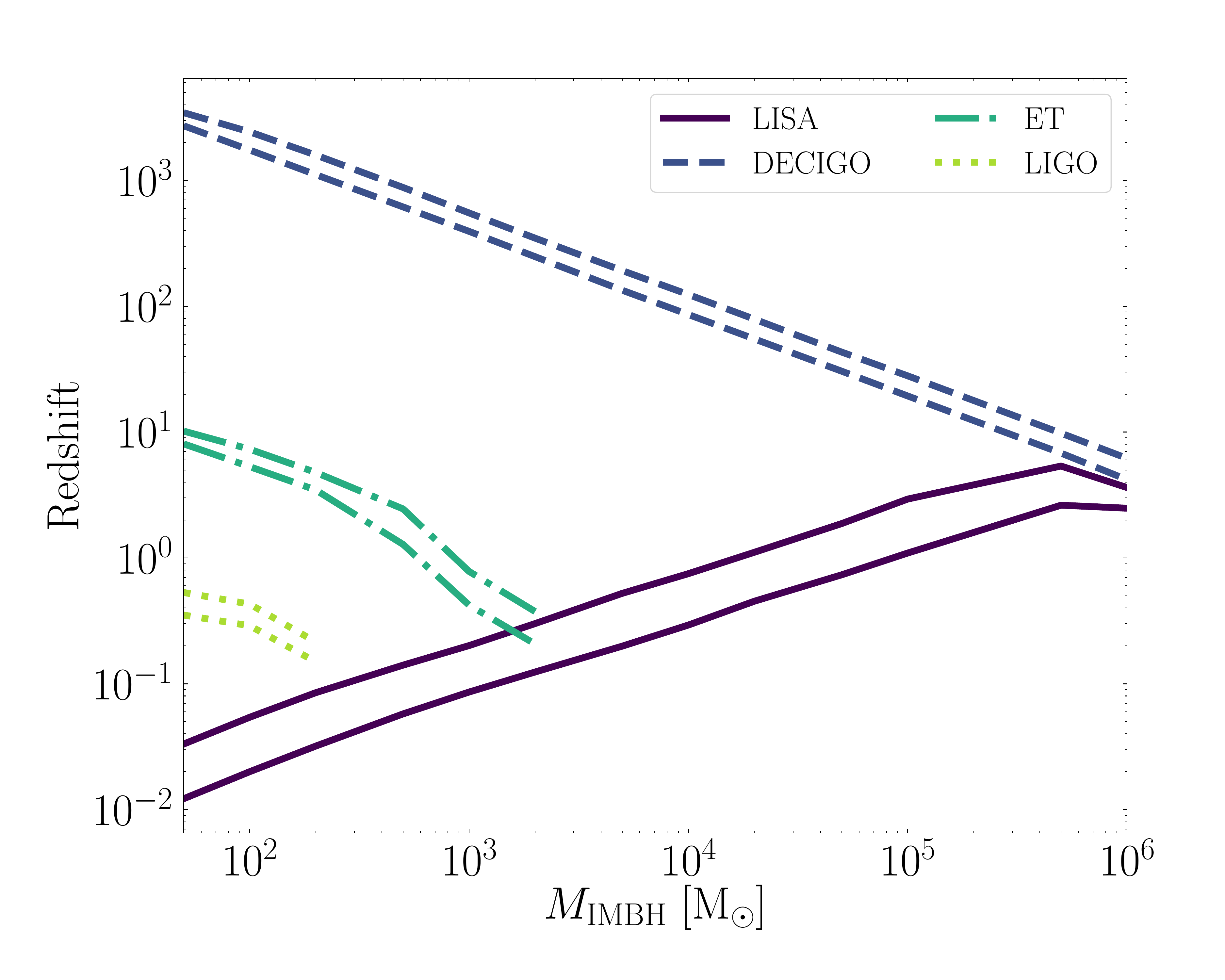}
\caption{Horizon redshift as a function of the IMBH mass assuming a BH companion with mass $10 \Ms$ (lower curves) or $30\Ms$ (upper curves). Different curve collections correspond to different detectors: LISA (straight lines), DECIGO (dashed lines), ET (dotted lines), LIGO (dot-dashed lines). In the next section, we discuss whether some classes of IMRIs can appear as multiband sources}.
\label{fig:fhor}
\end{figure}

To make predictions on IMRIs detectability, in this section we infer the merger rate for different detectors. To perform such calculations we need to account for the variation of galaxies number density across different redshifts $z$.
The GW source {\it horizon} determines the maximum distance in space, or the redshift $z_{\rm hor}$, at which the source signal is detected with a threshold signal-to-noise ratio (SNR), namely:
\begin{equation}
{\rm SNR^2} = \int_{f_1}^{f_2} \displaystyle{\frac{h_c^2(f,z_{\rm hor})}{S_n^2(f)}} {\rm d}f,
\end{equation}
with $f_{1,2}$ the initial and final frequency of the GW signal, $h_c(f,z)$ its characteristic strain, and $S_n(f)$ its sensitivity. To determine $z_{\rm hor}$ we integrate the final stage of the IMRI signal assuming an observation time of 4 yr -- i.e. the nominal duration time of the LISA mission -- and we assume an SNR of 15. We adopt the set of cosmological parameters measured by the Planck mission, namely $H_0 = 67.74$ km/s/Mpc$^{3}$, $\Omega_m = 0.3089$, $\Omega_\Lambda = 0.6911$ \citep{planck15}. Moreover, we vary the IMBH mass in the range $50-10^6\Ms$ assuming that the companion has a mass of either $10$ or $30\Ms$. Figure \ref{fig:fhor} shows how the horizon redshift changes for four different detectors: LIGO\footnote{https://www.ligo.caltech.edu/}, LISA\footnote{\url{https://www.elisascience.org/}}, DECIGO\footnote{\url{http://tamago.mtk.nao.ac.jp/spacetime/decigo_e.html}}, and ET\footnote{\url{http://www.et-gw.eu/}}.

The plot makes evident that ground based telescopes can provide insights on low-mass IMRIs ($<500\Ms$) up to redshift $z_{\rm hor} \leq 0.2$. GW190521, a GW source recently detected by the LIGO-Virgo collaboration \citep{gw190521a, gw190521b}, could be one of such low-mass IMRIs, composed of a stellar BH with mass $M_\bh = 16_{-3}^{+33}\Ms$ and an IMBH with mass $M_\ibh = 168_{-61}^{+15}\Ms$ \citep[see][]{nitz21}. The ET will enable the observation of IMRIs with mass $<10^3\Ms$ up to $z_{\rm hor} = 1-10$, whereas LISA will detect  IMRIs with  $M_{\ibh}=(10^4-10^5)\Ms$ in the same redshift. As such, the potential synergy between ET and LISA could enable GW astronomers to fully cover the whole range of IMBH masses. Decihertz observatories like DECIGO \citep{Kawamura11} and similarly designed mission \citep{arca19} will push the observational limits beyond LISA and ET, enabling a full coverage of IMBH mass spectrum up to the dawn of the Universe, thus constituting ideal detectors to unveil the truly nature of IMBHs. 

However, in the scenario explored here we assume that IMBHs form in GCs, thus the maximum redshift at which IMBH could be visible depends intrinsically on the typical timescales of star cluster formation. In the following, we adopt a maximum value of $z_{\rm max}=6$, corresponding to the formation epoch of the first stars, whenever $z_{\rm hor} > z_{\rm max}$. 

Once that the dependence between the horizon redshift and the IMBH mass is determined, we can estimate the total number of IMRIs inside the cosmological volume encompassed by the horizon, $N_{\imri}(z_{\rm hor})\equiv N_\imri$, a quantity that can be used to infer the IMRI merger rate. The $N_\imri$ parameter can be expressed as:
\begin{align}
N_{\imri} = & \Omega_s \int_{M_{1}}^{M_{2}} \int_{0}^{z_{\rm hor}} \frac{\derd n_{\imri}}{\derd M_\ibh\derd z} \times \nonumber \\
& \times \frac{\derd V_c}{\derd z}  \frac{\derd z}{1+z} \derd M_\ibh,
\end{align}
being $\derd V_c/\derd z$ the comoving cosmological volume element, $(1+z)^{-1}$ the term that account for the dilation time, and $\derd n_{\imri}/\derd M_\ibh$ the number of IMRIs per unit of IMBH mass.

The latter term depends intrinsically on the interplay among galaxies, star clusters, and IMBHs distribution within the cosmological volume. In the following, we pursue three different approaches to evaluate $\derd n_{\imri}/\derd M_\ibh$ and $N_\imri$, each implementing different assumptions about the distribution of IMBH hosts. In particular, the first approach (hereafter GAL) relies on the observed mass distribution of galaxies up to redshift $z=8$ \citep{conselice16}, the second approach (hereafter CFR) relies on the assumption that the star cluster formation rate (CFR) is nearly constant in the range $z=2-8$ and drops to zero at lower redshifts \citep{katz13}, while the third approach (CSFE) relies upon the cosmic star formation rate derived by \cite{madau17} and the cluster formation efficiency estimated by \cite{bastian08}.

In GAL we define 
\begin{align}
\frac{\derd n_{\imri}}{\derd M_\ibh} =& \xi_\bh f_{\gw} p_\ibh n_{\rm rep}\times \nonumber \\
& \times \frac{\derd n}{\derd M_g\derd z}\frac{\derd n_\gc}{\derd M_\gc}\frac{\derd M_\gc}{\derd M_\ibh}.
\end{align}
Here, $\xi_\bh$ represents the probability for the IMBH to form a binary with a stellar BH, $\derd n/(\derd M_g \derd z)$ represents the number of galaxies per unit of redshift and galaxy mass, $\derd n/\derd M_\gc$ is the number of clusters per cluster mass in a given galaxy, $\derd M_\gc/\derd M_\ibh$ connects GCs and IMBHs, $n_{\rm rep}$ is the number of times that the same IMBH can form an IMRI with a stellar companion, $f_{\gw}$ is the fraction of IMRIs that undergo merger within a Hubble time (a quantity that is extracted from our simulations, see Figure \ref{fig:gwprob}), and $p_\ibh$ represents the probability for a cluster to host an IMBH. In the following, we assume $p_\ibh = 0.2$ \citep{giersz15}. 

In a typical ensemble of stars with masses following a \cite{kroupa01} initial mass function, the number of stellar BH progenitors is a fraction $\sim 10^{-3}$ of the whole population. In absence of mass segregation and a mass spectrum, we might expect that the probability for an IMBH to be paired with a BH should simply be $10^{-3}$. However, in real systems mass-segregated stellar BHs tend to prevent other stars to migrate into the innermost cluster regions and thus they inhibit the IMBH to capture other stellar types. The direct implication of the dominant effect of stellar BHs on dynamics of the central cluster regions is a high probability for the IMBH to engage a long-term relationship with a stellar BH rather than with a star, thus suggesting $\xi_\bh \rightarrow 1$.

The term $\derd M_\gc / \derd M_\ibh$ can be calculated by inverting Equation \ref{mcmibh} and performing the derivative. For the number distribution of the GCs mass in a given galaxy, $\derd n/\derd M_\gc$, we assume a power-law
\begin{equation}
\frac{\derd n}{\derd M_\gc} = k M_\gc^{-s},
\end{equation}
with the slope $s = 2.2$\footnote{Note that this is compatible with the expected initial mass function of young and old star clusters in galaxies \citep[e.g.][]{gieles09}.} and the normalization constant 
\begin{equation*}
k = \frac{\delta M_g (2-s)}{(M_{\gc 2}^{2-s} - M_{\gc 1}^{2-s})}.
\end{equation*}
Assuming Galactic values for the galaxy stellar mass, $M_g = 6\times 10^{10}\Ms$, and star clusters mass range, $M_{\gc 1,2} = (5\times 10^3 - 8\times 10^6) \Ms$, we calculate the corresponding IMBH mass range $M_\ibh \simeq (30 - 4.6\times 10^4)\Ms$ according to Equation \ref{mcmibh}, which can also be used to write $M_\gc^{-s} = aM_\ibh^{-bs}$\footnote{The parameters $a,b$ are obtained manipulating Equation \ref{mcmibh}.}. The $\derd n/(\derd M_g\derd z)$ is obtained exploiting the results in \cite{conselice16}, who studied the distribution of galaxies with stellar masses up to $10^{12}\Ms$ up to redshift $z=8$. In particular, we exploit the following parametric expression of galaxies number density 
\begin{equation}
\phi(z) = -\frac{\phi_* 10^{(\alpha_* + 1)(M_2-M_*)}}{\alpha_* + 1},
\end{equation}
with $\phi_*,~\alpha_*,~M_*$ depending on the redshift \citep[see Table 1 in][]{conselice16}, and $M_2 = 12$. 

Substituting all the terms and manipulating them conveniently, the total number of IMRI in the portion of Universe accessible to a given detector is thus given by
\begin{align}
N_{\rm GAL} = & k a^{1-s} b p_\ibh n_{\rm rep} \xi_\bh \times \nonumber \\
& \times \int_{M_{1}}^{M_{2}} \int_0^{z_{hor}} f_{\gw} M_\ibh^{(1-s)b-1}  \derd M_\ibh \times \nonumber \\
& \times   \frac{\phi(z)}{1+z} \frac{\derd V_c}{\derd z}\derd z .
\label{eq:nimri1}
\end{align}

In both CFR and CSFE approaches, instead, we exploit the cosmological GC star formation rate $\rho_{\rm SFR}(z)$, which can be used to calculate the total number of GCs at a given redshift
\begin{equation}
N(z_{\rm max}) =  \int_0^{z_{\rm max}} \frac{\rho_{\rm SFR}(z)t(z)}{<M_\gc>}\frac{\derd V_c}{\derd z}\frac{\derd z}{1+z}.
\end{equation} 
Note that $t(z)$ expresses the dependence between time and redshift, and that the quantity $\rho_{\rm SFR}(z)t(z)/<M_\gc>$ represents the total number of GCs formed within redshift $z$.

Given the power-law GCs mass function used in the previous method, the normalization factor in this case become
\begin{align}
k  =&\frac{(1-s)}{M_{\gc, 1}^{1-s} - M_{\gc ,2}^{1-s}},
\end{align}
and the total number of IMRI inside a given cosmological volume is thus given by
\begin{align}
N_{\rm CFR, CSFE} = & k a^{1-s} b p_\ibh n_{\rm rep} \int_{M_{1}}^{M_{2}} \int_0^{z_{hor}} M_\ibh^{(1-s)b-1} \times \nonumber \\
& f_{\gw}  \rho_{\rm SFR}(z)t(z)\frac{\derd V_c}{\derd z} \frac{\derd z}{1+z} \derd M_\ibh.
\label{eq:nimri2}
\end{align}

In CFR, we assume that $\rho_{\rm SFR} = 0.005~\Ms~{\rm yr}^{-1}~{\rm Mpc}^{-3}$ in the range $z=2-6$ \cite[see][]{katz13}, whereas in CSFE we assume that GCs form following the cosmic star formation rate derived by \citep{madau17} with an efficiency $\eta_{\rm GC} = 0.08$ \citep{bastian08}, i.e.
\begin{equation}
\rho_{\rm SFR}(z) = \eta_{\rm GC}\psi(z), 
\label{CSFE}
\end{equation}
with
\begin{equation}
\psi(z)= \frac{0.01(1+z)^{2.6}}{1+[(1+z)/3.2]^{6.2}} ~\Ms~{\rm yr}^{-1}~{\rm Mpc}^{-3}.
\end{equation} 

The corresponding merger rate can be calculated as
\begin{equation}
\Gamma_{\imri} = \frac{N_\imri}{\langle T \rangle},
\end{equation} 
where $\langle T \rangle$ represents the median value of the IMRIs delay time, i.e. the time elapsed from the GC formation to the IMRI coalescence. 
By definition, for a typical IMRI $T$ is given by the sum of the cluster formation time $t_{\gc ,f}$, the IMBH formation time $t_{\ibh ,f}$, and the IMRI merger time $t_{\rm mer}$.

To estimate $\langle T \rangle$, we sample 2000 values of the triplet $(t_{\gc ,f}, t_{\ibh ,f},t_{\rm mer})$ as follows. 

The GC formation time $t_{\gc,f}$ is extracted according to either the GC formation rate derived by \cite{katz13} or the cosmic formation rate derived by \citep{madau17}. 
For the IMBH formation time $t_{\ibh, f}$, we assume that $1/3$ of IMBHs form via the rapid formation scenario whereas the remaining form via the slow scenario described in \citep{giersz15}. We thus extract $1/3$ of $t_{\ibh, f}$ values according to a uniform distribution limited within $0.05-1$ Gyr, and the remaining in the range $1-10$ Gyr. The IMRI merger time, instead, is sampled directly from our simulations (model S0). 

According to the procedure above, the median value of the IMRIs delay time is ${\rm Log} \langle T \rangle = 9.14 \pm 1.22$. 

\begin{figure}
\centering
\includegraphics[width=\columnwidth]{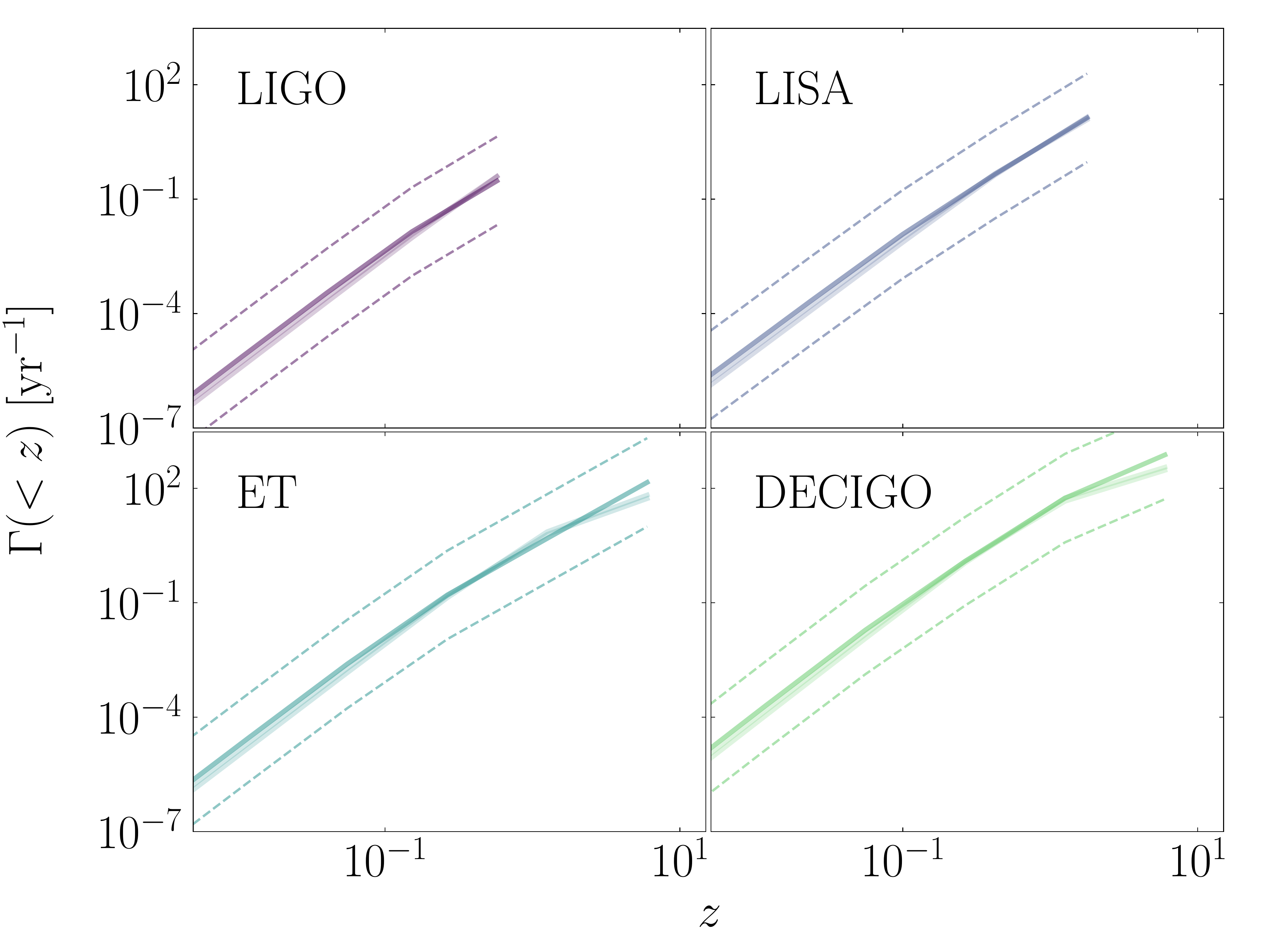}
\caption{Cumulative merger rate as a function of redshift for different detectors. Thick lines represent the results from GAL model, whereas the thin lines correspond to CFR and CSFE, assuming the mean value of the elapsed time. The dotted lines represent the upper(lower) bounds as determined by the minimum(maximum) value of $\langle T \rangle$. We assume that the BH companion in the IMRI has a mass $M_\bh = 30\Ms$.}
\label{fig:rate}
\end{figure}

Figure \ref{fig:rate} shows the IMRI merger rate as a function of redshift calculated for different approaches, different $\langle T \rangle$ values, and for different instruments (LIGO, LISA, ET, and DECIGO).

The high uncertainties in $\langle T \rangle$ affect significantly our estimates. 

Assuming a $T_{\rm obs} = 4$ yr long observation run, we find for LIGO a total merger rate $\Gamma_\imri T_{\rm obs} \sim 0.08 - 24$ yr$^{-1}$ out to a redshift $z_{\rm hor, max} = 0.57$. 

At lower frequencies, LISA might record up to $\Gamma_\imri\sim 0.8-200$ yr$^{-1}$, pushing the limit for the IMBH mass to up to $40,000\Ms$ and thus allowing to explore the mass range typical of IMRIs $q \simeq 10^{-4}$. 

While the constraints on "present-day" technologies are already quite encouraging, the next generation of both ground- and spaced-observatories could enable us to deliver a much larger amount of observations up to the epoch of the formation of the first stars, thus allowing us to probe different IMBH formation mechanisms.

For instance, the ET could detected up to $\Gamma_\imri \sim 4-2000$ yr$^{-1}$ with masses $M_\imri < 2,000 \Ms$, while space observatories sensitive at decihertz frequencies like DECIGO might record up to $\Gamma_\imri \sim 20-10^4$ yr$^{-1}$.

Table \ref{tab:3} summarises the average IMRI rate values for the range of approaches and detectors adopted. 

\begin{table*}
\caption{Main properties of our models}
\begin{center}
\begin{tabular}{ccccccccccc}
\hline 
\hline
Instrument & $M_{\rm SBH}$ & $z_{\rm max}$ & $M_{\ibh , max}$ & $\Gamma_1$ & $\Gamma_2$ & $\Gamma_3$ & $\Delta\Gamma_1$ & $\Delta\Gamma_2$ & $\Delta\Gamma_3$ \\
   & $\Ms$ &  & $\Ms$ & yr$^{-1}$ & yr$^{-1}$ & yr$^{-1}$ &yr$^{-1}$ & yr$^{-1}$ & yr$^{-1}$ \\ 
\hline
LIGO & $10$ & $0.38$ & $200$     & $0.04$  &$0.04$ &$0.04$  &$0.003-0.54$  &$0.003-0.59$  &$0.003-0.54 $ \\
LIGO & $30$ & $0.57$ & $200$     & $0.09$  &$0.11$ &$0.11$  &$0.006-1.3 $  &$0.008-1.6$   &$0.008-1.6  $ \\
LISA & $10$ & $0.70$ & $46240$   & $0.35$  &$0.45$ &$0.45$  &$0.024-5.1 $  &$0.031-6.5$   &$0.031-6.5  $ \\
LISA & $30$ & $1.78$ & $46240$   & $3.9$   &$3.5$  &$4.1$   &$0.27 -56.2$  &$0.24 -50.0$  &$0.28 -59.3 $ \\
ET & $10$ & $6.00$ & $2000$      & $27.7$  &$10.9$ &$13.5$  &$1.9  -399.7$ &$0.75 -157.2$ &$0.94 -195.5$ \\
ET & $30$ & $6.00$ & $2000$      & $41.3$  &$15.5$ &$19.2$  &$2.8  -596.5$ &$1.1  -224.0$ &$1.3  -278.1$ \\
DECIGO & $10$ & $6.00$ & $46240$ & $217.2$ &$85.5$ &$103.2$ &$15.0 -3139$  &$5.9  -1235$  &$7.1  -1492 $ \\
DECIGO & $30$ & $6.00$ & $46240$ & $217.2$ &$85.5$ &$103.2$ &$15.0 -3139$  &$5.9  -1235$  &$7.1  -1492 $ \\
\hline
\end{tabular}
\tablefoot{Col 1: instrument name. Col 2: mass of the IMRI secondary. Col 3: horizon redshift. Col 4: maximum IMBH mass visible. Col 5-7: average merger rate obtained with models GAL, CFR, CSFE. Col 8-9: boundaries of the inferred merger rate for models GAL, CFR, CSFE.}
\end{center}
\label{tab:3}
\end{table*}

\section{Conclusions}
\label{end}

In this work we explored one possible mechanism for the formation of IMRIs in dense globular clusters, namely the chaotic interaction among two BHs and one IMBHs. Using a large set of $N$-body models, we studied how different properties (stellar BH mass spectrum, host cluster mass, IMBH mass) affect IMRIs development. We exploited these models to investigate the possible implications for IMBH seeding and growth in dense clusters, and to infer the potential implications for present and future ground- and space-based GW detections. 
Our main results are summarised in the following.

\begin{itemize}
\item The probability for IMRIs formation and merger is noticeable, $P_{\rm mer} = 5-50\%$, and maximizes at larger values of the IMBH mass $M_\ibh$. 
\item The mass distribution of stellar BHs involved in merging IMRIs maps out the underlying stellar BH mass spectrum, thus suggesting that IMRIs are possible probes of the BH mass spectrum in dense clusters. A statistically significant number of IMRIs could tell us whether the BH mass spectrum in star clusters containing an IMBH preserves the original shape or is shaped by other mechanisms, e.g. the aforementioned BH burning via dynamical interactions. 
\item Interestingly, this formation mechanism leads to IMRIs having generally low average eccentricities ($e = 10^{-4} - 0.02$) when sweeping through the GW frequency bands typical of low- ($10^{-3}-10^{-1}$ Hz), intermediate- ($10^{-1}-1$ Hz) and high- ($1-100$ Hz) detectors. The low eccentricity could be indicator of this specific formation channel, as other scenarios (Kozai-Lidov resonant systems, hyperbolic encounters, gravitational captures) are expected to produce a consistent amount of eccentric IMRIs.
\item We couple the results of the $N$-body models with a semi-analytic tool to explore the probability for IMRI remnants to be retained in their host clusters. We find that, due to GW recoil, only IMBH with masses above $M_\ibh = 10^3\Ms$ have a significant retention probability ($>75\%$), provided that their companion BH mass is lighter than $M_\bh < 45\Ms$. For lighter IMBHs ($M_\ibh \sim 10^2\Ms$) we derive a retention fraction smaller than $10\%$, regardless of the BH spin distribution and the IMBH spin. 
\item For IMBHs with masses $M_\ibh < 10^3\Ms$, we show that even a single merger leaves an imprint on the spin of the remnant. In the case of Schwarzschild IMBHs, a merger with a stellar BH could increase the remnant spin to up to $S_\ibh = 0.7(0.2)$ for IMBHs with a mass $M_\ibh = 10^2(10^3)\Ms$. Similarly, for nearly extremal IMBHs the spin after one single merger can decrease down to $S_\ibh = 0.8(0.2)$ in the same IMBH mass range.
\item The IMBH spin evolution is particularly interesting in the case of multiple mergers. We show that IMBHs growing via multiple mergers should exhibit low-spins, generally $S_\ibh < 0.2$. Thus, detecting IMRIs occurring in a star cluster would give us crucial insights on IMBHs formation mechanisms: IMBHs formed via long merger chains would preferentially have small spins, whilst those formed mostly via stellar feeding and collapse of massive stars are likely to follow the BH natal spin distribution. 
\item We explore whether the IMRI formation channel discussed in this work could sustain the IMBH growth in dense globular or nuclear clusters. Using a semi-analytic tool to model IMBH evolution, we show that IMBH seeds heavier than $M_{\rm seed} > 300\Ms$ can grow up to $M_{\ibh} >10^3\Ms$ via multiple mergers. Assuming that such seeds forms at redshift $z\sim 2-6$, we predict that around $1-5\%$ of them would reach typical masses $\sim 500-1500\Ms$ at redshift $z=0$ in massive globular clusters.
\item We show that LISA can detect IMRIs in Milky Way globular clusters with a signal-to-noise ratio (SNR) up to SNR$=10-100$, and in Large Magellanic Cloud clusters with an SNR$=8-40$.
\item We derive IMRIs merger rate for several detectors: LIGO ($\Gamma_{\rm LIGO} = 0.003-1.6$ yr$^{-1}$), LISA ($\Gamma_{\rm LISA} = 0.02-60$ yr$^{-1}$), ET ($\Gamma_{\rm ET} = 1-600$ yr$^{-1}$), and DECIGO ($\Gamma_{\rm DECIGO} = 6-3000$ yr$^{-1}$). Our estimates highlights how the future synergy among GW detectors would enable us to fully cover the mass range bridging stellar BHs and SMBHs.
\end{itemize}

\section*{Acknowledgements}

MAS acknowledges the Alexander von Humboldt Foundation for the financial support provided in the framework of the research program "The evolution of black holes from stellar to galactic scales", the Volkswagen Foundation Trilateral Partnership project No. I/97778 ``Dynamical Mechanisms of Accretion in Galactic Nuclei'', and the Sonderforschungsbereich SFB 881 "The Milky Way System"  -- Project-ID 138713538 -- funded by the German Research Foundation (DFG).  PAS acknowledges support
from the Ram{\'o}n y Cajal Programme of the Ministry of Economy, Industry and
Competitiveness of Spain.  This work was supported by the National Key R\&D
Program of China (2016YFA0400702) and the National Science Foundation of China
(11873022, 11991053).  The authors acknowledge support from the COST Action
GWverse CA16104. The authors acknowledge the use of the Kepler computer at ARI
Heidelberg, funded by Volkswagen Foundation through the project GRACE 2:
"Scientific simulations using programmable hardware" (VW grants I84678/84680),
and the bwForCluster of the Baden-W\"urttemberg's High Performance Computing
(HPC) facilities, which is supported by the state of Baden-Württemberg through
bwHPC and the German Research Foundation (DFG) through grant INST 35/1134-1
FUGG.

\bibliographystyle{aa}
\footnotesize{
\bibliography{ASetal2015}
}

\end{document}